\documentclass[journal]{IEEEtran}

\usepackage{xcolor,soul,framed} 
\usepackage{xcolor}

\colorlet{shadecolor}{yellow}
\usepackage[pdftex]{graphicx}
\graphicspath{{../pdf/}{../jpeg/}}
\DeclareGraphicsExtensions{.pdf,.jpeg,.png}

\usepackage[cmex10]{amsmath}
\usepackage{array}
\usepackage{mdwmath}
\usepackage{mdwtab}
\usepackage{eqparbox}
\usepackage{url}
\usepackage{algorithm,algpseudocode,amsmath}
\usepackage{float}
\usepackage{xspace}
\usepackage{amssymb}
\usepackage{multicol}
\usepackage{subfigure}
\usepackage{multirow}
\usepackage{url}

\usepackage[sorting=none]{biblatex}

\usepackage{biblatex}
\begin{document}
\bstctlcite{IEEEexample:BSTcontrol}
    \title{Dual Protection Ring: User Profiling Via Differential Privacy and Service Dissemination Through Private Information Retrieval}
    \author{
        \textsuperscript{[1]}~Imdad Ullah, \textsuperscript{[2]}~Najm Hassan, \textsuperscript{[3]}~Tariq Ahamed Ahangar, \\
         \textsuperscript{[4]}~Zawar Hussain Shah, \textsuperscript{[5]}~Mehregan Mahdavi
         \textsuperscript{[6]}~Andrew Levula\\
        {\small
            \textsuperscript{[1]}~School of Computer Science, Faculty of Engineering, The University of Sydney, Sydney NSW 2006, Australia. 
             \textsuperscript{[2]}~Higher Colleges of Technology, United Arab Emirates (UAE).
            \textsuperscript{[3}~Management Information Systems Department, College of Business Administration, Prince Sattam bin Abdulaziz University, Al-Kharj 16278, Saudi Arabia. 
            \textsuperscript{[4]}~Department of Information Technology, Sydney International School of Technology and Commerce, Sydney NSW 2000, Australia.
            \textsuperscript{[5]}~Kingsford Institute of Higher Education (KIHE), Sydney NSW 2000, Australia.
            \textsuperscript{[6]}~School of Business, Excelsia College, Sydney NSW 2113, Australia.         
            \\
            \vspace{2mm}
            \textsuperscript{[1]}~imdad.ullah@sydney.edu.au
             \textsuperscript{[2]}~nhassan@hct.ac.ae
             \textsuperscript{[3]}~t.ahanger@psau.edu.sa,
            \textsuperscript{[4]}~zawar.s@sistc.nsw.edu.au
            \textsuperscript{[5]}~dean@kihe.com.au,
            \textsuperscript{[6]}~andrew.levula@excelsia.edu.au,
        } 
    }

\maketitle

\begin{abstract}
User profiling is crucial in providing personalised services, as it relies on analysing user behaviour and preferences to deliver targeted services. This approach enhances user experience and promotes heightened engagement. Nevertheless, user profiling also gives rise to noteworthy privacy considerations due to the extensive tracking and monitoring of personal data, potentially leading to surveillance or identity theft. We propose a dual-ring protection mechanism to protect user privacy by examining various threats to user privacy, such as behavioural attacks, profiling fingerprinting and monitoring, profile perturbation, etc., both on the user and service provider sides. We develop user profiles that contain sensitive private attributes and an equivalent profile based on differential privacy for evaluating personalised services. We determine the entropy of the resultant profiles during each update to protect profiling attributes and invoke various processes, such as data evaporation, to artificially increase entropy or destroy private profiling attributes. Furthermore, we use different variants of private information retrieval (PIR) to retrieve personalised services against differentially private profiles. We implement critical components of the proposed model via a proof-of-concept mobile app to demonstrate its applicability over a specific case study of advertising services, which can be generalised to other services. Our experimental results show that the observed processing delays with different PIR schemes are similar to the current advertising systems.
\end{abstract}



\begin{IEEEkeywords}
Privacy, Private information retrieval, Information leakage, Privacy threats, Tracking, User profiling

\end{IEEEkeywords}

\IEEEpeerreviewmaketitle



\section{Introduction} \label{section-introduction}
\sloppy User profiling is a valuable tool for delivering personalised services and is the practice of collecting and analysing user data for their interests, behaviours, and preferences in different domains ~\cite{kanwal2024machine, wang2024know, muller2024big}. Profiling can improve user experience by reducing irrelevant interests to ensure that the user is merely presented with relevant and engaging content.  
User profiling's evolution from stereotype models to deep learning, along with the necessity of precise user representations derived from implicit data and multi-behavior modeling (incorporating graph structures), has been thoroughly examined, with an emphasis on privacy, explainability, and fairness \cite{purificato2024user}.
Service-oriented companies use this information to provide personalized recommendations and services tailored to the user's needs and preferences, e.g., an e-commerce website may use profiling to suggest products \cite{hu2023effect}\cite{rehman2022impact} or advertising systems to target users with personalised ads \cite{truong2019integrated}\cite{jebarajakirthy2021mobile}\cite{pooranian2021online}\cite{ullah2020privacy}. Although user profiling can provide several benefits, it exposes sensitive information that users may not be aware of, e.g., tracking a user's search/browsing history, location data, personal information, political beliefs, sexual orientation etc. 
In this direction, \cite{maraj2024survey} presents a comprehensive survey that addresses user profiling, data collection, and the associated privacy issues prevalent in contemporary Internet services.  
Hence, if data are not properly protected, it can be accessed by malicious actors, leading to privacy violations, identity theft and tracking \cite{jiang2024pervasive}, discrimination, and other harm. In addition, user profiling can be used to create extremely engaging and addictive content to keep users obsessed with online content for longer periods, which may lead to a potentially negative impact on one's mental health and general well-being.  
To ensure user privacy, a two-pronged approach is necessary, encompassing the perspectives of both users and the measures taken by online service providers to safeguard users against targeting. Hence, safeguarding individual interests resulting from user profiling is a crucial aspect of protecting user privacy alongside the measures taken by service providers to safeguard users' privacy. There are several techniques to preserve online user privacy, such as, browser extension based on crypto techniques \cite{guha2011privad, DBLP:conf/ndss/ToubianaNBNB10, ullah2023privacy, backes2012obliviad, zhong2022ibex, liu2020security}, user-defined selection of private attributes \cite{sanchez2018privacy, ullah2014profileguard}, techniques based on anonymity \cite{checco2022opennym}\cite{bussard2004untraceable}, randomisation \cite{quoc2017privacy} \cite{kim2021successive}, and other techniques based on differential privacy \cite{wang2018toward} \cite{chen2023differential} and private information retrieval \cite{mozaffari2020heterogeneous} \cite{bodur2023private}. It is worth noting that most existing solutions do not offer private systems that ensure privacy protection from the users' and servers' perspectives.  

This paper introduces a dual-ring protection mechanism to safeguard user privacy on both the user and service provider sides when providing and retrieving services. We examine the various privacy threats users face, such as profiling based on opt-in services, user interactions with service providers, and app usage activity. Attackers may exploit users by revealing their privacy and carrying out attacks such as behavioural attacks, private information usage, monitoring attacks, profile fingerprinting, stealing sensitive information, and profile perturbation. Similarly, service providers may also exploit user privacy by exposing private attributes, stealing and selling sensitive information, exploiting the integrity of profiling attributes, carrying out perturbation, and presenting targeted services. To mitigate these privacy attacks, various measures can be taken on the user end, such as limiting the magnitude of private attributes in the user profile, protecting attributes during profile updates, allowing users to privately interact with servers, securing sensitive attributes, and privately interacting with service providers. Similarly, service providers can also be restricted to ensure that user privacy is not exploited during service dissemination, collection of sensitive attributes, profiling in various groups, and maintaining information consistency and service integrity.

In this study, we examine the profiling process, which involves various entities collecting sensitive attributes, such as analytics companies, to target users with personalised services. This process entails gathering private sensitive attributes, such as user identifiers, demographic information, behavioural data, and data obtained by manipulating analytical insights. Subsequently, we create an equivalent profile based on differential privacy that can be used to evaluate personalised services within a service marketplace. This differential privacy mechanism is also applied when the user profile is updated to ensure that sensitive attributes are not exposed to service marketplaces or third-party attackers. To protect sensitive profiling attributes, we further evaluate the entropy of the resultant profile, where we either employ data evaporation by distorting controlled data to increase entropy or destroy private profiling attributes artificially. Additionally, on the server side within the service marketplace, we use a private information retrieval mechanism to evaluate personalised services against the differentially private profile. The resultant services are retrieved via private information retrieval, which may contain a comprehensive set of relevant objects provided by the service marketplace or retrieved from a third-party. Therefore, our approach ensures private interaction with third-party services, including analytics services. Finally, we evaluate the impact of our proposed scheme on a specific case study involving advertising services, where users are targeted with advertisements in mobile apps based on their private and sensitive profiling attributes. 


Our proposed scheme involves several changes on the user side within a service marketplace ecosystem. Users are provided with a series of functionalities, such as locally evaluating and updating their profile to reflect various activities, sending their profile for differential privacy evaluation, and encoding and decoding queries to retrieve relevant services from the service marketplace privately. Additionally, we implemented a Proof of Concept on an Android-based device to implement private information retrieval and interact with desktop-based server machines and the profiling process. This implementation required significant changes to run private information retrieval schemes since the CPU architecture of Android-based devices is entirely different from desktop machines. Our Proof-of-Concept implementation aimed to demonstrate the applicability of our proposed scheme in a real-world scenario. 


We conducted two experimental setups as part of our study. The first setup had two sub-experiments: one for profiling, where we generated comprehensive sets of profiling attributes over 10,800 hours, and another for ads collection, where we collected ads over a total app run of 2,160 hours. We selected different apps from various contexts, such as Business or Sports. We evaluated different classes of received ads, including those related to the user profile, context, generic, and random ads. We comprehensively evaluated the impact of differential privacy on various statistics related to different classes of ads across various context profiles. We analysed the percentage increase/decrease of these ads, their distribution over time, the impression and idle time, and the distribution of the unique ads among different context profiles and frequency bins. Similarly, our second experimental setup was to prove the applicability of private information retrieval (PIR). To mimic the real advertising ecosystem, we set up at least three PIR servers. We varied these servers from 3 to 6, along with other necessary settings for running various PIR schemes, such as the number of colluding servers. We run these experiments over information-theoretic and hybrid PIR schemes. The query request was encoded with varying ad requests against the server databases of up to 10GB. We implemented these schemes on our Proof-of-Concept client, which encoded the queries with the tested PIR schemes. We observed that the processing delays with different PIR schemes were similar to those in the current advertising ecosystem.

Our main contributions to this work are: 
\begin{itemize}
    \item To propose a protection mechanism to safeguard user privacy on both the user and service provider sides when providing and retrieving services. 
    \item To investigate the profiling process involving various entities collecting sensitive attributes by analytics companies to target users with personalised services.
    \item A model that provides insights into the establishment of Interest profiles by individual apps in the Context profiles and then shows how the profiles evolve when some apps other than the initial set of apps are utilised.
    \item To evaluate the entropy of the profile and employ data evaporation by distorting controlled data to increase entropy or destroy private profiling attributes artificially.
    \item To evaluate the impact of our proposed scheme on a specific case study involving advertising services, where users are targeted with advertisements in mobile apps based on their private and sensitive profiling attributes.
    \item Experimental setups for generating comprehensive sets of profiling attributes over 10,800 hours and for ads collection with over a total app run of 2,160 hours. 
    \item Experimental setup for running PIR with multiple servers to mimic the real-world advertising system.
    \item Implement a Proof-of-Concept on an Android-based device with private information retrieval that interacts with desktop-based server machines and the profiling process.
\end{itemize}


The paper is structured as follows: Section \ref{section-related} provides an overview of the related work. Section \ref{section-problem} describes the research problem and the proposed model. Section \ref{system-components} introduces the components of the proposed system. Section \ref{proposed-scheme} outlines our proposed scheme, including the differential privacy-based profiling process and the private information retrieval for personalised service retrieval. Section \ref{performance-measures} discusses the performance measures used in our evaluation. Section \ref{evaluations} presents the experimental evaluations, including the experimental setups used. In Section \ref{discussion}, we discuss our proposed scheme. Finally, we conclude the paper in Section \ref{conclusion}. 

\section{Related work} \label{section-related}

Online service providers are increasingly collecting personal information, building user profiles, and monetizing them with advertisers and data brokers, leaving users with little control over their personal information. In a recent study \cite{kaifa2023demystifying}, 458 third-party libraries in 642 Android-based apps were investigated, revealing that 23\% of the libraries violate regulation requirements for providing privacy policies, 37\% do not disclose users' data usage in their libraries, and 52\% of apps share data with third-party libraries. The authors in \cite{cooper2023privacy} analyzed the usefulness of programmatic advertising in various aspects, such as information collection and data collection on consumer publications, and they found that 26\% of consumers in the US are driven by strong preferences for ads relevant to their behavior over the internet, and 34\% of users show a strong desire to prevent advertisers from aggressively collecting their personal data. Similarly, \cite{nunez2023role} designed a survey-based study conducted on over 24,000 internet users to study the usefulness, annoyance, privacy concerns, and their concerns over data storage and private information usage via cookies.
\cite{ciuchita2023programmatic} surveyed programmatic advertising in online retail to understand consumer perceptions and data-sharing risks involving automated tools. Other studies also examine online consumer information sharing, including \cite{martin2017data}, \cite{yan2023subscriptions}, \cite{chen2023double}, \cite{gao2023information}, and  \cite{lukic2023impact}.


According to \cite{li2017resolving}, inconsistencies exist between users' stated preferences and their actual online behavior, even though users generally prefer to disclose less information \cite{taddicken2014privacy}. Privacy is recognized as a multi-dimensional concept that requires attention from all stakeholders \cite{belanger2011privacy}. A psychological framework has been proposed to identify various dimensions of users' privacy concerns \cite{stuart2019psychology}. A privacy-preserving framework called "Gist" uses an aggregate model to balance user privacy with data utility and was evaluated using a dataset of 0.1 million users from the census bureau \cite{bilogrevic2014s}.


\sloppy The previous works can be broadly classified into two categories: those that focus on minimising the data shared with third parties and those that rely on anonymising proxies to protect user behavioural data from online trackers until users agree to sell their data. Examples of the former include local user profiling techniques \cite{akkus2012non, guha2011privad, mohan2013prefetching, toubiana2010adnostic}, while examples of the latter include \cite{backes2012obliviad, riederer2011sale}. Another recent work \cite{du2022optimal} presents a privacy-preserving targeted advertising strategy for virtual reality applications that communicate via edge access points. The authors derive an advertising strategy based on the Vidale-Wolfe model and Hamiltonian function, using input budget to boost advertising revenue and preserve user privacy. Additionally, \cite{Chen:2013:Splitx, chen2012towards} propose privacy-preserving data aggregation using a combination of homomorphic encryption and differential privacy. However, these proposals suffer from relying on third parties, intensive use of public key for each user input or XOR encryption, and linkability of user information to individual users upon decryption.


It is important to note that there is a limited number of works that address user privacy from both the user and advertising system sides while incorporating user profiling based on differential privacy. While there are several works that implement differential privacy for preserving privacy in other environments, such as smart metering for renewable energy \cite{hassan2019differential}, privacy preservation in building data \cite{janghyun2022review}, privacy in wireless sensor networks \cite{zhang2023dp}, and PrivSTD \cite{wang2021privacy}, which is an edge computing-based privacy-preserving mechanism for streaming crowdsourced data. Our approach to protecting user privacy involves several steps, including evaluating profiling attributes locally, calculating their differentially private version, and privately computing the queries for encoding and decoding communication that take place between the user and service marketplace.

\section{Research Problem} \label{section-problem}
Advertising and analytics companies collect individuals' sensitive data, which users may benefit from, by presenting them with personalised content and advertising. However, it leads to potential harm, including privacy violations, selling or disseminating sensitive data, or manipulation. The collected data can involve a rich set of data, such as browsing history, location data, social media activities, and data related to race, gender, or other protected characteristics; furthermore, using deep learning techniques may come up with additional dimensions of sensitive data. This information, if not adequately protected and used, can be accessed by malicious users, leading to identity theft/fraud, or used by companies to profile and target users with personalised services, or discriminatory practices based on race, gender etc., or such information could be used to create engaging or addictive technologies to harm youngsters particularly. 

\subsection{User profiling}
The advertising and analytics companies profile users using sophisticated techniques that collect user behavioural data, analyse, and create detailed profiles that reflect the individual's interests, behaviour, and preferences. These techniques include tracking cookies, device and browser fingerprinting, social media data, location data, user surveys, and surrounding people's behaviour (based on location and friends and family in the circle) and their interests. Similarly, for the in-app ads and services, the advertising and analytics companies use tracking libraries to track user activities over the `Web \& App'\footnote{E.g.., Activity controls with Google: \url{https://myactivity.google.com/activitycontrols?utm_source=my-activity}} that include, app usage behaviour, web browsing histories, and data collected from the partnering tracking companies. This information is used to profile users and is then used to target them with personalised content e.g., advertisement services customised to individual/group of users. 

In addition, these companies may contain additional information, such as demographics (i.e., age, gender), (an example user profile with Google\footnote{My Google activity: \url{https://myactivity.google.com/myactivity}}) that are represented as the profiling interests, which we call \textit{Interest profiles}. Furthermore, the \textit{Context profile} is characterised as the mobile apps installed from various categories defined by the services platforms; a detailed discussion over these profiles is given in Section \ref{representing-user-profile}. Usually, these profiles are created due to the user's interactions with various services, e.g., Google uses AdMob SDK \cite{ullah2014profileguard} tracking libraries installed on user devices or attached to individual apps that communicate with various entities, e.g., Analytics servers, to derive individual's interest. The tracking data can be exported in different formats using Google Takeout\footnote{\url{https://takeout.google.com/}}, which contains a wide range of data, including email conversations, calendars, browsing histories, etc.

\subsection{Entities in action: Collecting Sensitive Information}
\begin{figure*}[h]
\begin{center}
\includegraphics[scale=0.4]{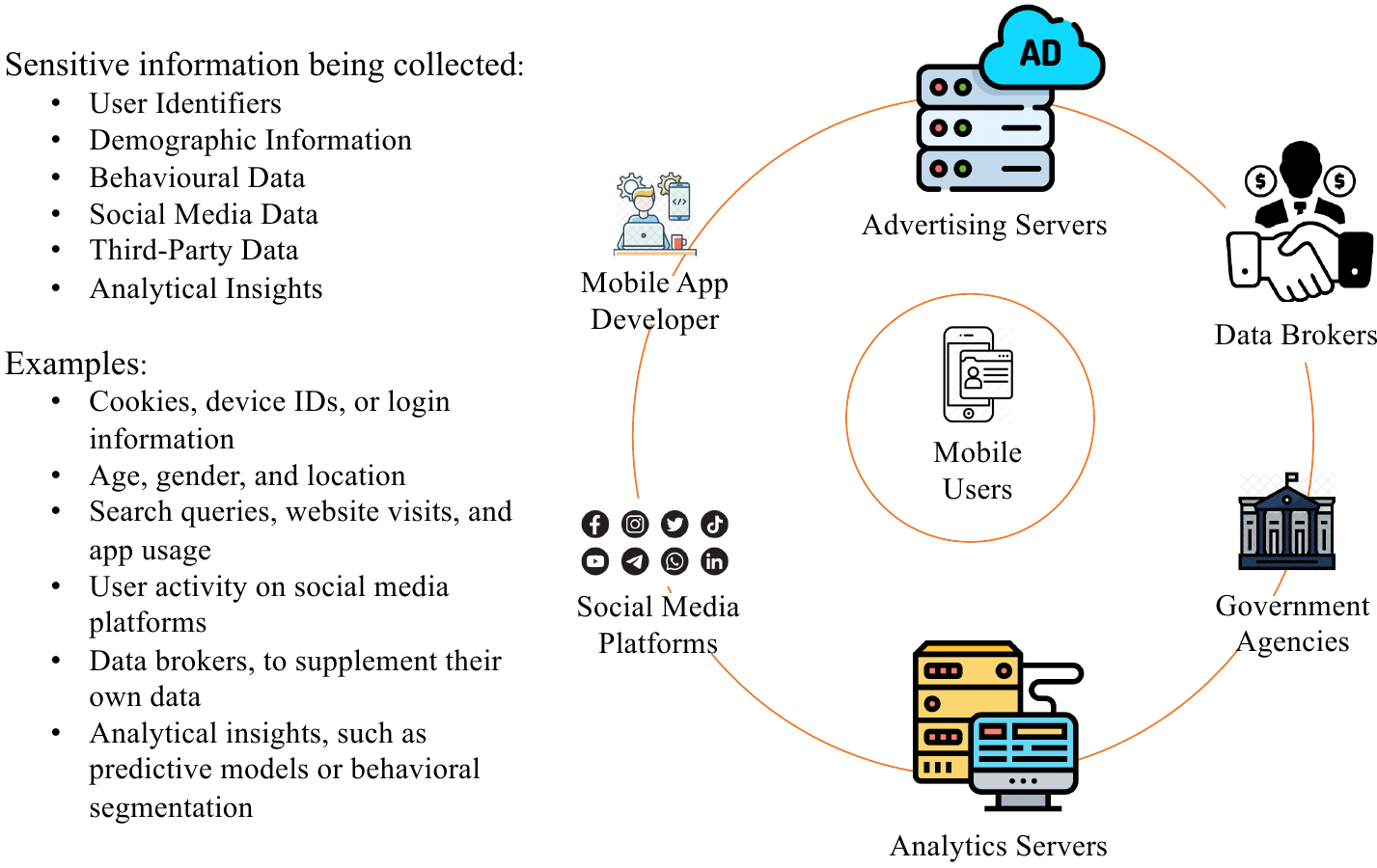}
\caption{Entities that collect sensitive information and comprehensive examples of the types of information that may be collected and exchanged.}
\label{collection-entities}
\end{center}
\end{figure*}

\sloppy Figure \ref{collection-entities} shows (not limited to) various entities that collect and profile users for sensitive information. These entities include \textit{mobile users} who use various products/apps and services and either provide their personal information or the companies infer this information based on their usage and overall behaviour. The \textit{advertising companies} that utilise user data to create and serve targeted advertisements; the \textit{analytics companies} that collect and analyse user data to provide insights into user behaviour, preferences, and trends; such insights can be used to improve products, services, and marketing strategies. Similarly, the \textit{data brokers} collect and sell user data to other companies that may collect data from multiple sources, such as public records, social media, and online tracking. The \textit{social media platforms} allow users to share personal information; they often collect and share user data with advertising and analytics companies. The \textit{government agencies} collect and use user data for law enforcement and national security purposes. In addition, the \textit{mobile app developers} that create mobile apps often implant analytics libraries that collect user data, including location data, contact information, and browsing history. It is important to note that there may be other entities involved in collecting and profiling user data, at the same time, the relationship among these entities may be complex and constantly evolving.

Figure \ref{collection-entities} also shows comprehensive examples of the types of information that may be collected and exchanged; see the left side of this figure. These data collecting entities may exchange a variety of sensitive information, which depends on the companies involved and the types and purpose of data being collected. This data is classified into profiling components that may be extensively complex and reflect users' interests, behaviours, and preferences, causing potential harm to users' privacy.

\subsection{Research Statement} \label{section-statement}
\begin{figure*}[h]
\begin{center}
\includegraphics[scale=0.4]{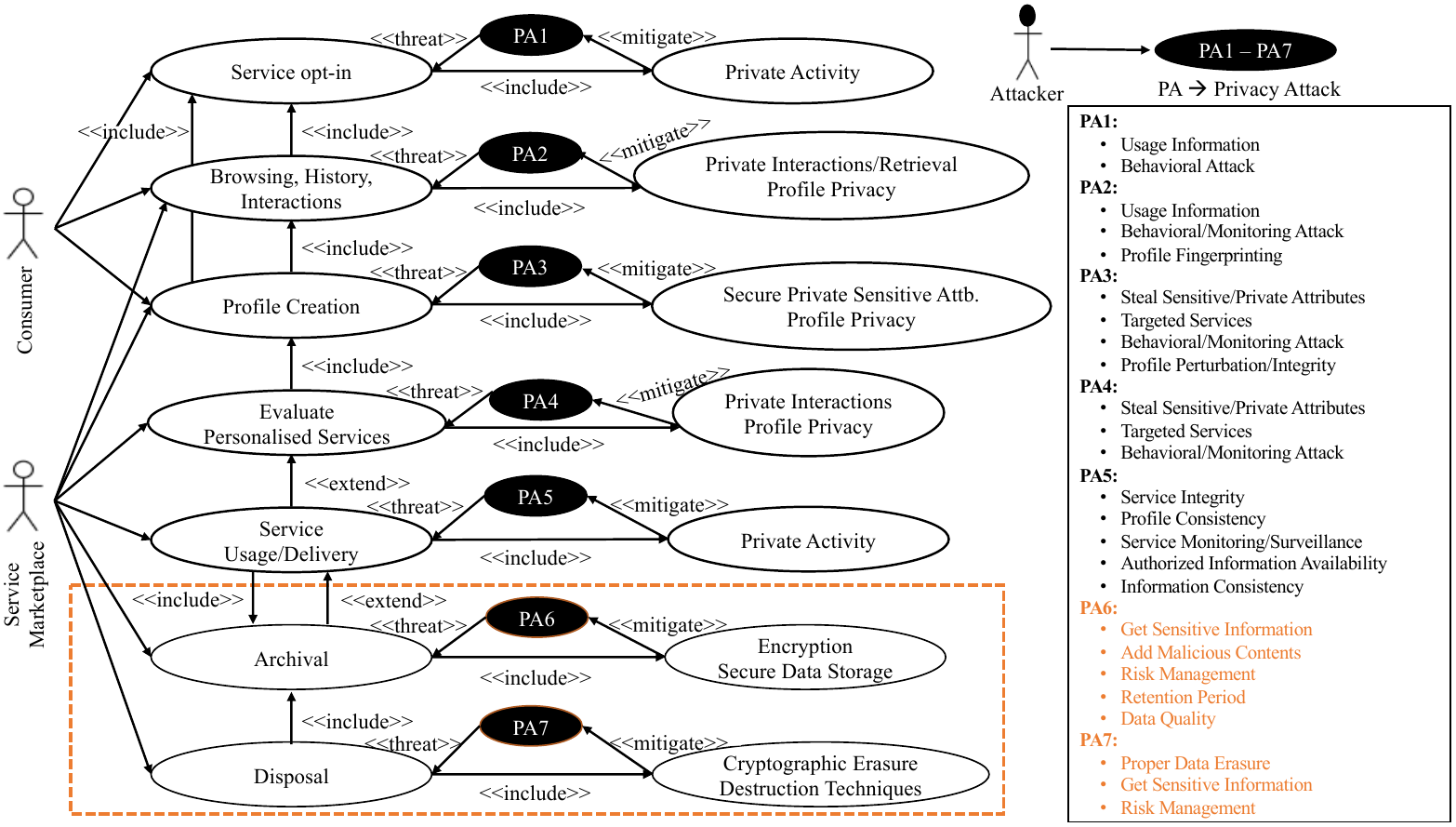}
\caption{Threat actor's misuse cases: Legitimate components of the service marketplace (ovals on the left), identified threats (PA1 to PA7), and mitigation approaches (ovals on the right). We plan to carry out a comprehensive study of orange-coloured enclosed segments (i.e., legitimate components, associated threats, and mitigation approaches) in future, which are essential parts of the risk assessment.}
\label{miuse-case}
\end{center}
\end{figure*}

Figure \ref{miuse-case} shows a comprehensive overview of threat actor's misuse cases showing legitimate portions of the service marketplace, threats to these portions, and how to mitigate identified threats effectively. Specifically, we show legitimate use cases of an advertising system (ovals to the left of the figure), privacy threats (shaded ovals, i.e. PA1 to PA7; their description is given to the right of the figure), shaded ovals connected to unshaded ones with an arrow labelled \texttt{<<threaten>>} to show threats to a specific portion of the system, unshaded ovals to show how to mitigate these misuses, i.e., connected to shaded ones with arrows labelled \texttt{<<mitigate>>}, in addition to, legitimate (e.g., consumer, service marketplace) and threat actors (stick figure with shaded head). For example, a threat (say PA3) threatens the Profile Creation (a legitimate portion of the system) process where an attacker is interested in various dishonest activities, such as ``Steal Private Attributes, Targeted Services, Behavioural Profile Attack, and Profile Integrity'', which can be mitigated via ``Secure Private/Sensitive Attributes, and Profile Privacy''. Note that we mainly include threats related to the private profiling process, resultant service dissemination to associated consumers, behavioural activities, and service usage profile. 

\subsection{Threat Model} \label{threat-model}
We aim to achieve user privacy during the profiling process and privately disseminate the services to the users. This eventually achieves privacy on the user side and the service providers/marketplaces. We begin by profiling users, including profiling interests and context information, then updating and continual interaction with the service marketplace and protecting those profiling attributes via differential privacy. Subsequently, these profiles are passed on to the marketplace, where personalised services are calculated via PIR and interacted with the consumers. We measure the randomness/uncertainty of the calculated profile before sending any sensitive profiling attributes to the marketplace, where, based on the threshold, various mechanisms are invoked to either distort or destroy the private profiling attributes. We assessed our proposed framework over the advertising service as a case study where we applied differential privacy to the user profiles and investigated its effect on the collected mobile advertisements. Furthermore, for the applicability of the PIR mechanism, we create a different setup to simulate an actual environment that consists of several PIR servers that interact with mobile devices in a real-time environment. 

\section{Proposed System Components} \label{system-components}

We formalise the system model that consists of the \textit{apps}' profiles, interests' profiles, and the conversion of resulting profiles by use of applications in an \textit{app} profile. In particular, we provide the insights of \textit{establishment} of \textit{Interests profiles} by individual \textit{apps} in the \textit{Context profiles} and then show how the profiles \textit{evolve} when some \textit{apps} other than the initial set of \textit{apps} are utilised.

\subsection{System Model}\label{system-model}

Service marketplaces utilise various personalisation and customisation technologies \cite{braynov2003personalization, tang2010combination} to infer user information to offer better products and services tailored to individual interests. These techniques include filtering technologies (Rule-based, Collaborative, and Content-based filtering), Web usage analysis, Intelligent agents, Location-based personalisation, and Preference modelling. These techniques associate specific personalised items with individual consumers, record their temporal activities, or make informed recommendations to users by finding a set of neighbouring users with interests similar to the target users. The authors in \cite{tang2010combination} use a systematic approach to extract and integrate interests from different resources and discover user profiling interests to construct a user interest model. 

Following, we denote a service by ${s_{i,j}}$, $i=1,...,S_j$, where $S_j$ is the number of subservience that belong to a service category $\Theta _j$, here $j=1,..., \theta $ and $\theta $ is the number of different service categories provided by any online services platform e.g., Amazon, Google, Apple etc. E.g., Google\footnote{\url{https://support.google.com/a/answer/182442?hl=en}} provides various services such as Google Workspace, Google Workspace Marketplace Apps, which can be personalised to various groups within an organisation. Additionally, an individual service category $j$ may have numerous services, which can be opt-in for a specific team within an organisation and used, e.g., the Marketing team could be customised to use Google Pay (i.e., a specific service $i$) service. We note that these individual services implement their own analytics and tracking capabilities to track the activities of individuals and groups, profile their behaviour and actions, and subsequently target them with private and personalised services. 

Moreover, we note that an individual service $s_{i,j}$ provides various facilities to carry out an enterprise's operational business processes. For example, Google Analytics is a web analytics service for its marketing platform that tracks users' activities, reports the website traffic, and tracks social networking sites and applications. Similarly, other activities, such as ``Web \& App Activity'' services of Google, record user's activities on its sites, apps, and locations to better facilitate faster searches, better recommendations, and more personalised experiences in Cloud, Maps, Search, Drive and other Google services. These various services profile individual users with their tracking capabilities and target them with personalised services, e.g., targeted services, personalised advertisements, etc. Hence, a user may be characterised by a set of individual services used by a particular consumer or a group of users. 

Let $\mathbb{S}_s \in \mathcal{S}$ represent a set of services a user opts in; $\mathcal{S}$ represents the total number of services a particular service platform provides. In particular, various services provided by Google, a $\mathbb{S}_s$ may consist of Google Analytics, Search Console, Domains, Cloud, Jamboard etc. We call this a \textit{context profile} ${K_s}$ that can be defined as:

\begin{equation}\label{initial-app-profile}
{K_s} = \left\{ {\left\{ {{s_{i,j}},{\Theta _j}} \right\}:{s_{i,j}} \in {\mathbb{S}_s}} \right\}
\end{equation}

The service-based platforms partially profile and target their associated consumers based on the combination of various services currently used and their consecutive interactions. To comprehensively represent such information within their personalised profile, we presume the following condition: It is vital to validate that the total number of services a user opts-in, i.e., $n\left( {{K_s}} \right)$, does not exceed an entire set of various services i.e., $0 < n\left( {{K_s}} \right) \le \mathcal{S} $. In addition, to effectively profile a consumer and to offer customised services, a user must opt-in to at least one service, i.e., $n\left( {{K_s}} \right) \ge 1$; ideally, a user is fundamentally tracked for mobile apps and browser activities through the ``Web \& App Activity'' services of Google. Furthermore, it is crucial to ensure that a particular subservice belongs to a specific set (i.e., ${\Theta _j}\left( {{s_{i,j}}} \right) \ne 0 $) and is not undefined within a service marketplace (i.e., $\theta \subseteq {\Theta _j}\,\,;\,\,\forall j$). Various important notations and their descriptions are presented in Table 1.

\subsection{User Interest Profiling}\label{representing-user-profile}
We represent profiling interest by  ${r_{k,l}}$ that is derived by the analytics capabilities of a specific service ${s_{i,j}}$; here $k=1,...,R_l$, where $R_l$ set of interests defined under the specific categories $\Omega_l$, $l=1,..., \omega $, $\omega $ is the set of interest categories defined by analytics companies/service marketplace. The set of profiling interests $\mathcal{R}$, i.e., the characteristics that may be assigned to users, are grouped under an \textit{interest profile}. Note that two services ${s_{i,j}}$ may derive identical interests, which shows a strong association of the user's interests to derived categories. An example user \textit{interest profile}\footnote{My Ad Center: \url{https://myadcenter.google.com/controls?hl=en}. Note that additional services can be added to personalise your interests further.} consists of various profiling interests derived from various services linked to the Google platform. The summary of services used can be browsed through Google Dashboard\footnote{\url{https://myaccount.google.com/dashboard?hl=en&pli=1}}. The profiling interests are typically grouped under various classified interests' categories with specific interests \cite{ullah2022joint} that are eventually used for \textit{service targeting}. Following, we represent an \textit{interest profile} $ I_r $ that is used to enable targeted services: 

\begin{equation}\label{eq:initial-interes-profile}
{I_r} = \left\{ {\left\{ {{r_{k,l}},{\Omega _l}} \right\}:{r_{k,l}} \in {\mathbb{S}_r}} \right\}
\end{equation}

\begin{figure}[h]
\begin{center}
\includegraphics[scale=0.31]{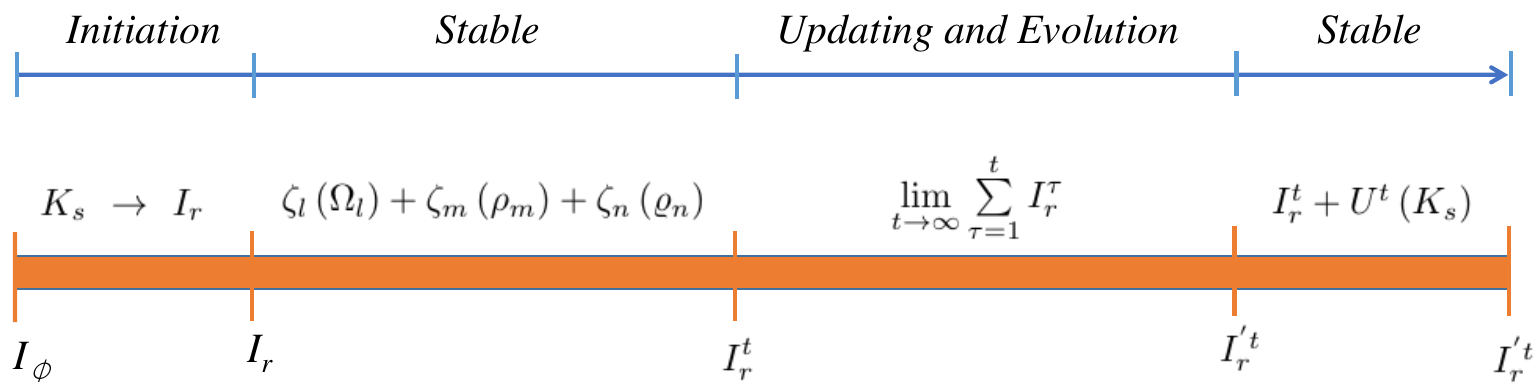}
\caption{Various stages during the profiling process: the profile \textit{initiation}  during \textit{iprofile establishment} stage, the \textit{stable} state with no impact of further activity, and the \textit{evolution} state that reflects overall user activities.}
\label{profile-development}
\end{center}
\end{figure}

It is important to note that an individual interest ${r_{k,l}}$ comprises a set of keywords \({\kappa _{k,l}}\), which characterises specific profiling interests along with its category. The $ I_r $ continuously undergo different processes to fully reflect the true interest of individuals, e.g., \textit{profile establishment} to initially derive $ I_r $ from the service usage i.e., \({K_s} \to {I_r}\) using a \textit{mapping} function $f$ i.e., \(f:{\Theta _j} \to {\Omega _l} = {\mathbb{S}_{{K_s}}}\left\{ {{\Omega _l}} \right\}_{l = 1}^\omega \)) i.e., to generate specific profiling interests \({S_{{K_s}}}\left\{ {{\Omega _l}} \right\}_{l = 1}^\varepsilon \), see \cite{ullah2014profileguard} for detailed discussion over \textit{profile establishment} process. Another process is the \textit{profile evolution} $ I{'_r}$ to capture variations in $ I_r $ observed over time. Similarly, the \textit{profile development} process $I_r^f$) i.e., $I_r^f = {I_r} \cup I{'_r}$ where the further activity of specific services does not have any further impact over a specified time. The profiling process is presented in Figure \ref{profile-development} that shows various stages, i.e., the profile \textit{initiation}  for initial profile establishment, the \textit{stable} state with no impact of further activity, and the \textit{evolution} state where various behavioural activities are reflected over time.

The interest categories $\Omega_l$ present in $I_r^f$ are present with a certain proportion based on which the \textit{service targeting}is carried out. Hence, it is important to determine the dominating interests categories and assign certain weights (i.e., \({\zeta _l}\left( {{\Omega _l}} \right)\)) along with other non-dominating categories based on following constraints: $\zeta _l^{\min } \le {\zeta _l}\left( {{\Omega _l}} \right) \le \zeta _l^{\max },\,\forall l \in {\Omega _l}|l \in \omega$ and $0 < {\zeta _l}\left( {{\Omega _l}} \right) \le \zeta _l^{\max } $. These two constraints respectively ensure that an interest category \({\zeta _l}\left( {{\Omega _l}} \right)\) is within the \(\zeta _l^{\min }\) and \(\zeta _l^{\max }\) threshold \textit{weightages} and non-negativity constraint. Subsequently, the user profile is represented as:

\begin{equation}
{I_r} = {\zeta _l}\left( {{\Omega _l}} \right),\forall l \in \omega 
\end{equation}

\begin{table*}[t]
\begin{center} \scalebox{0.8}{
\label{table:notations}
{\begin{tabular}{c|l}
\hline
 \textbf{Symbols}& \textbf{Description}\tabularnewline
\hline
${\mathcal{S}}$& Available \textit{services} in a marketplace \tabularnewline
\hline
$\theta$ & Total number of \textit{service} categories, $\Theta_j$ is a selected category, $j=1,...,\theta$ \tabularnewline
\hline
 ${\mathbb{S}_s}$& Subset of \textit{service} opt-in on a user's device\tabularnewline
\hline
 ${s_{i, j}}$& A \textit{service} ${{s_{i,j}} \in S}$, $i=1,...,S_j$, $j=1,...,\theta$, $S_j$ is the number of \textit{service} in $\Theta_j$\tabularnewline
\hline
\({\kappa _{i,j}}\)& Set of keywords associated with individual \textit{service} ${s_{i,j}}$ including its category\tabularnewline
\hline
 $K_s$& Context profile consisting of \textit{service} $s_{i,j}$ and their categories $\Theta_j$ \tabularnewline
\hline
 $\mathcal{R}$& Set of interests in marketplace's interests list\tabularnewline
\hline
 $\Omega_l$& Interest category in $\mathcal{R}$, $l=1,...,\omega$, $\omega$ is the number of interest categories defined by a marketplace \tabularnewline
\hline
 $S_r$& Subset of $\mathcal{R}$ derived by $S_s$\tabularnewline
\hline
 $I_r$& Interest profile consisting of ${r_{k,l}}$, ${r_{k,l}} \in S_r$\tabularnewline
\hline
 ${r_{k,l}}$& An interest in $I_r$, $k \in R_l$, $l \in \omega$ \tabularnewline
\hline
\({\kappa _{k,l}}\)& Set of keywords associated with individual interest ${r_{k,l}}$ including interest category\tabularnewline
\hline
 $\mathbb{S}_r^{i,j}$& Set of interests derived by an \textit{service} $s_{i,j}$ \tabularnewline
\hline
$f$ & Mapping that returns the derivation of \textit{services'} (or history, ad's category) category to interest category \tabularnewline
\hline
\({\zeta _l}\left( {{\Omega _l}} \right)\)& The \textit{weightage} assigned to interest category in $\Omega _l$ in a profile with, respectively with \(\zeta _l^{\min }\) and \(\zeta _l^{\max }\) thresholds \tabularnewline
\hline
\({{\rho _m}}\)& Profiling components for representing \textit{browsing history/searches} \tabularnewline
\hline
\({{\varrho _n}}\)& Profiling components for representing \textit{interactions with ads} \tabularnewline
\hline
\({C_t}\left( {{\zeta _l}\left( {{\Omega _l}} \right)} \right)\)& Tracking changes in interest category due to \textit{changes in services'} usage\tabularnewline
\hline
\({C_t}\left( {{\zeta _m}\left( {{\rho _m}} \right)} \right)\)& Tracking \textit{change in web browsing history/searches}\tabularnewline
\hline
\({C_t}\left( {{\zeta _n}\left( {{\varrho _n}} \right)} \right)\)& Tracking \textit{interactions with ads} i.e., clicks\tabularnewline
\hline
\({U{'^\tau }\left( {{K_s}} \right)}\) & Usage of \textit{services} in $K_s$ at time slot $t$; average usage of each \textit{service} is given by \({\overline {U{'^\tau }\left( {{K_s}} \right)} }\)\tabularnewline
\hline
\({\zeta _{l'}}\left( {{\Omega _{l'}}} \right)\) & The \textit{weightage} assigned to interest category generated by recommended service(s), where \(l' \ne l\) \tabularnewline
\hline
$\mathcal{D}$& The set of rows in $\mathcal{D}$ that consists of $D=N_{r,s},I_{r}^{\tau}$, $N_{r,s}$ i.e., specific interests and $r$ and opt-in services $s$ along with $I_{r}^{\tau}$ \tabularnewline
\hline
$\mathcal{D}$&A database that contains profiling interests from users of a marketplace \tabularnewline
\hline
$\mathcal{DB}$ & A database that maintains relevant data related to services $s_{i,j}$ such as, description, keywords, etc. \tabularnewline
\hline
$\beta^{th}$ & A specific record that a user is interested to privately retrieving via PIR from a database server $\mathcal{DB}_{n}$ \tabularnewline
\hline
$ L\left(I_{r},t\right)$& Privacy loss at time $t$ when a subset of attributes of $I_r$ are disclosed \tabularnewline
\hline
$ H\left(I_{r},t\right)$& Entropy level of private attributes in $I_r$. Maximum entropy level is represented as $ H^{*}\left(I_{r}\right)$  \tabularnewline
\hline
$\mathcal{C}$& The set of categories defined by mapping platforms, such as URL-Classification\footnote{\url{https://url-classification.io/}}, with $c$ a category from top to root-level\tabularnewline
\hline
${d_{r,s,q}}$ &  An ad in the set of ads ${\mathbb{D}_{r,s}}$, $r$ is specific user profile, $s$ is service's category, and $q$ is the numbered ad \tabularnewline
\hline
$f$ & Mapping that returns an either direct mapping of the ads category or its keywords ${\kappa_{r,s,q}}$ within $\mathcal{C}$\tabularnewline
\hline
 ${E_{r,s}}$& Set of ads' with along with their transformed categories $c$ \tabularnewline
\hline
$\delta _p^{ran}$& The set of \textit{random} ads for particular profile $p \in P$. Represented as $\delta _{r,s}^{tar}$ \textit{targeted}, $\delta _{r,s}^{con}$ \textit{contextual}, and $\delta _{r,s}^{gen}$ \textit{generic} ads\tabularnewline
\hline
$C_s$, $C_r$ &Respectively as the sets of transformed services and profiling interest's categories \tabularnewline
\hline
\end{tabular}} }
 \end{center} \caption{List of Notations}
 
\end{table*}


\subsection{Service Browsing History, Searches, and Service Interactions}\label{browsing-searches}

A consumer is partly profiled for their activities, which are recorded by services analytics, e.g., how frequently a user browses or searches for a specific service or the frequency of interactions. We represent the \({\rho _m}\) and \({\varrho _n}\) as the profiling components, respectively, for browsing or searching and interactions with a particular service. The assigned weightages of both these components are: $\left( {\zeta _l^{\min },0} \right) < {\zeta _m}\left( {{\rho _m}} \right) \le \zeta _l^{\max },\,\,\forall m,\forall l \in \left[ {1,\omega } \right]$ and $\left( {\zeta _l^{\min },0} \right) < {\zeta _n}\left( {{\varrho _n}} \right) \le \zeta _l^{\max }\,\,\forall n,\forall l \in \left[ {1,\omega } \right] $. These constraints ensure the assignment of higher weightage to both \({\rho _m}\) and \({\varrho _n}\) components to show more significance of the preference of a user. Hence, the resultant profile becomes:

\begin{equation}\label{eq:profile-weightages}
{I_r} = {\zeta _l}\left( {{\Omega _l}} \right) + {\zeta _m}\left( {{\rho _m}} \right) + {\zeta _n}\left( {{\varrho _n}} \right)
\end{equation}

\subsection{Profile Updating and Evolution}\label{profile-updating}

The profiling updates incorporate user activities and temporal changes in a user profile, and hence, \textit{service targeting}, which changes the targeting criteria with time. There are different measures to record such changes in a profile, e.g., a user opts in/out or increases/decreases interactions. These changes evolve with time $t$, represented as \({C_t}\left( {{\zeta _l}\left( {{\Omega _l}} \right)} \right)\); \({C_t}\left( {{\zeta _m}\left( {{\rho _m}} \right)} \right)\) and \({C_t}\left( {{\zeta _n}\left( {{\varrho _n}} \right)} \right)\) respectively for changes in service usage, browsing through particular services, and the frequency of service interactions. Hence, Eq. (\ref{eq:profile-weightages}) can be rewritten as $ I_e^t = {C_t}\left( {{\zeta _l}\left( {{\Omega _l}} \right)} \right) + {C_t}\left( {{\zeta _m}\left( {{\rho _m}} \right)} \right) + {C_t}\left( {{\zeta _n}\left( {{\varrho _n}} \right)} \right)$. It is important to ensure the non-negativity constraint as well as the maximum bounding thresholds for the three interest types i.e., $0 < {C_t}\left( {{\zeta _l}\left( {{\Omega _l}} \right)} \right) \le C_t^{\max }\left( {{\zeta _l}\left( {{\Omega _l}} \right)} \right);\,\,\forall t,\forall l \in \left[ {1,\omega } \right] $, and $0 < {C_t}\left( {{\zeta _m}\left( {{\rho _m}} \right)} \right) \le C_t^{\max }\left( {{\zeta _m}\left( {{\rho _m}} \right)} \right);\,\,\forall t $, and $0 < {C_t}\left( {{\zeta _n}\left( {{\varrho _n}} \right)} \right) \le C_t^{\max }\left( {{\zeta _n}\left( {{\varrho _n}} \right)} \right);\,\,\forall t $.

These components are also mapped to profiling interests \({\Omega _l}\) so as to produce a unified user profile i.e., \(f:{\varrho _n} \to {\Omega _l} = {S_{{K_s}}}\left\{ {{\Omega _l}} \right\}_{l = 1}^\omega \). These newly evolved changes are deterministically bounded by a finite constant, e.g., \(C_t^{\max }\left( {{\zeta _m}\left( {{\rho _m}} \right)} \right)\), and distributed with an unknown probability distribution. Hence, the user profile evolves as a unified profiling interest, as follows:

\begin{equation}\label{eq:profile-final}
I_r^t = \sum\limits_{\forall t} {{C_t}\left( {{\zeta _l}\left( {{\Omega _l}} \right)} \right)}; \,\,\,\forall l \in \left\{ {\left[ {1,\omega } \right],m,n} \right\}
\end{equation}

Furthermore, at time $t+1$ the profiling interests evolves i.e.,  \(I_r^{t + 1} = I_r^t + {C_{t+1}}\left( {{\zeta _l}\left( {{\Omega _l}} \right)} \right)\). Hence, for $n$ amount of time: 

\begin{equation}\label{eq:profile-change-final}
0 < I_r^{t + 1} \le I_r^{t + n}\,\,\,\,\forall t
\end{equation}

Here, the \(I_r^{t + n}\) is the maximum convergence point of an interest profile; similarly as \({t \to \infty }\) the profile gains maximum  convergence i.e., $\mathop {\lim }\limits_{t \to \infty } \sum\limits_{\tau = 1}^t {I_r^\tau }$. The maximum profiling convergence highly reflects the dominance of consumers' private attributes when targeted with personalised services. Hence, our goal is to fetch user profiles over user devices via differential privacy, calculate thresholds via entropy, invoke a particular mechanism for reducing the overall entropy of a user profile, and request the services using private information retrieval. Furthermore, we are particularly interested in business operational interactions (e.g., uploading private content to servers, interactions with servers for profiling purposes, requesting particular objects such as targeted advertisements or objects from content delivery networks, etc.) where a request is privately fulfilled against a diverse range of servers. 

\section{Proposed Scheme}\label{proposed-scheme}
\begin{figure}[h]
\begin{center}
\includegraphics[scale=0.3]{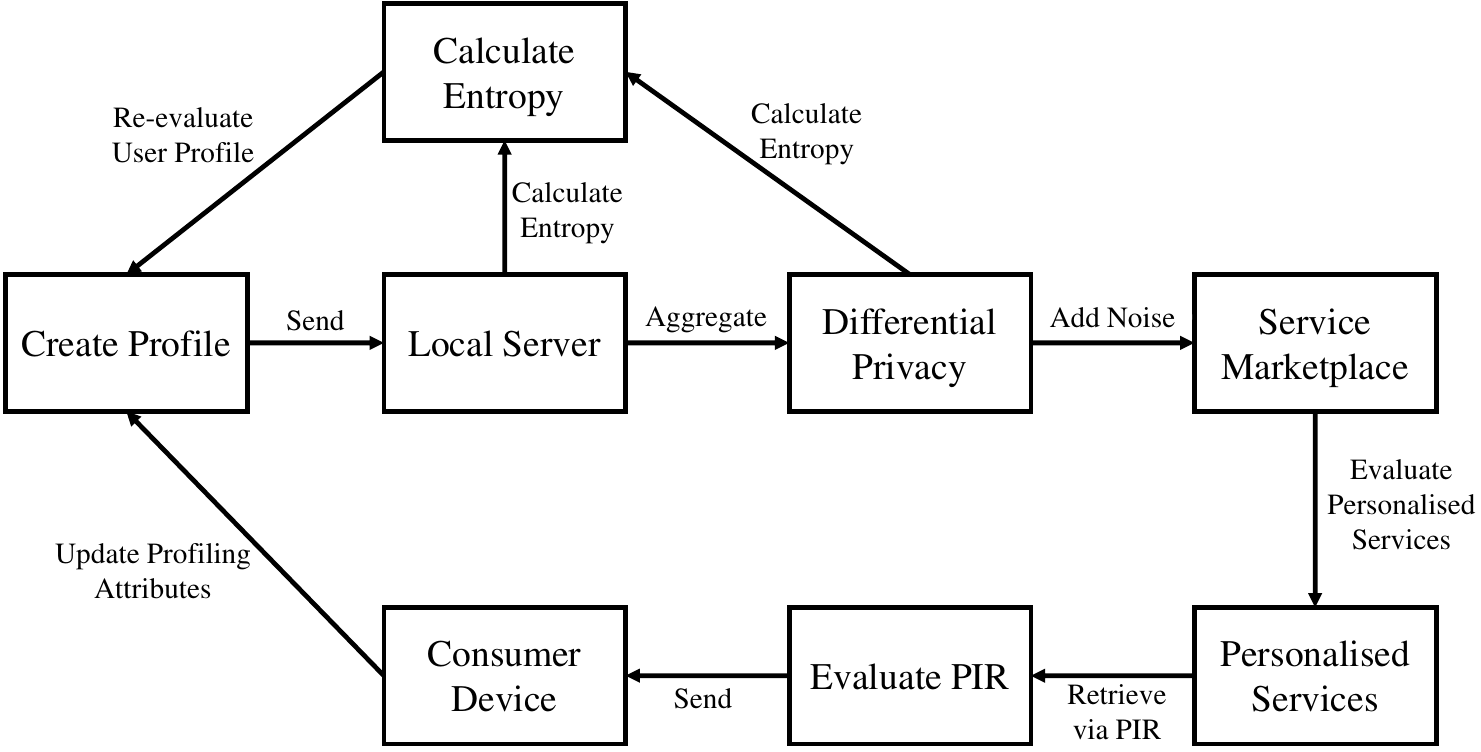}
\caption{The locally created user profile is sent to LS, which calculates differentially private profile that is used to evaluate personalised services within the Service Marketplace via PIR. Entropy is calculated to evaluate the disclosure of sensitive attributes.}
\label{system-process}
\end{center}
\end{figure}

Figure \ref{system-process} shows the various steps of the proposed scheme: the user locally creates behavioral user profile that represents various private profiling attributes (i.e., Sections \ref{representing-user-profile}, \ref{browsing-searches}, \ref{profile-updating}, and \ref{service-usage-profiling}) and sends it to Local Server (LS); the LS evaluates equivalent differentially private profile (Section \ref{profiling-diff}), which is used to evaluate personalised services within the Service Marketplace (Section \ref{personalised-services}). These services are retrieved via PIR (Section \ref{viaPIR}) and service interactions takes place within the Consumer Device, which further contributes to evaluating profiling interests based on user interactions. We also propose to evaluate entropy of the profiling attributes to evaluate disclosure of sensitive profiling attributes at LS (and once the differential privacy is evaluated), which might lead to re-evaluating the profiling interests (Section \ref{evaluate-entropy}).  

\subsection{Profiling Interests via Differential Privacy}\label{profiling-diff}
The consumer sends Interest profile to the Local Server before sending it to the service marketplace, where it locally evaluates the differential privacy for profiling interests of several users. This makes the service marketplace difficult to determine the contribution of individual consumers by adding noised to the data. Let $D=N_{r,s},I_{r}^{\tau}$ denotes the set of rows in a database $\mathcal{D}$ that consists of $N_{r,s}$ representing specific interests $r$ and the opt-in service(s) $s$ along with profiling interests $ I_{r}^{\tau}$ received from various users. Note that specific information, such as personal identifiers, are not sent to the Local Server. We generalise the Local Server that can be specific service (e.g., Maps) where the consumer does not want to directly send profiling information to the Local Server, however it should locally perform the private information process. Consequently, the consumer decides to use services without involving an intermediate entity and to make it easier while interacting for frequently exhaustive operations. 

The Local Server utilises randomised algorithm $\mathcal{M}$ over $\mathcal{D}$ and generates the output $\mathbf{o}$ i.e., $\mathcal{M:\mathcal{D\rightarrow\mathcal{O}}}$ that satisfies the $\epsilon$-differential privacy conditioned to $O\subseteq\mathcal{O}$. Following:

\begin{equation}
Pr\left[\mathcal{M}\left(D\right)\in O\right]\leq e^{\epsilon}\cdot Pr\left[\mathcal{M}\left(D'\right)\in O\right]
\end{equation}

The $\mathcal{M}$ satisfies the $\epsilon$-differential privacy with guaranteeing stronger privacy for smaller value of $\epsilon$, conditioned to changing the attributes on an individual record in $\mathcal{D}$ that has negligible effect on the output $O$ of $\mathcal{M}$. 

In order to calculate statistical information as a query performed over $\mathcal{D}$, such as the most visited/requested service, it is important to calculate the notion of sensitivity of the output. Let $\mathbf{Q}:D\rightarrow\mathbb{N^{d}}$ is a query that evaluates most requested service; $d$ is the number of elements in output. Following, let the $\mathbf{Q\left(D\right)}$ and $\mathbf{Q}\left(D'\right)$ are two $d$-dimensional vectors, then the sensitivity of $\mathbf{Q}$ can be calculated as $\Delta\left(\mathbf{Q}\right)=max_{D,D\prime\in D}\parallel\mathbf{Q\left(D\right)}-\mathbf{Q\left(D'\right)}\parallel$, where $\parallel\mathbf{Q\left(D\right)}-\mathbf{Q\left(D'\right)}\parallel$ be the norm of $\mathbf{Q\left(D\right)} $ and $\mathbf{Q\left(D'\right)} $ (i.e., distance from the origin) for all neighboring $D,D'\in\mathcal{D}$.  

A common differential privacy method ads Laplace noise to the output $\mathbf{o}$ using Laplace Perturbation Algorithm (LPA) \cite{dwork2011firm, dwork2006our}. The output $\mathbf{o}$ via $\mathcal{M}$ by adding independently generated noise form a zero-mean Laplace distribution with scale parameter $\lambda=\Delta\left(\mathbf{Q}\right)/\epsilon$ to each $d$ output values of  $\mathbf{Q}$ \cite{dwork2011firm}:

\begin{equation}\label{diff-output}
\mathbf{o}=\mathbf{c}+\left[Lap\left(\ensuremath{\lambda=\Delta\left(\mathbf{Q}\right)/\epsilon}\right)\right]^{d}
\end{equation}

Eq. \ref{diff-output} achieves the $\epsilon$-differential privacy. Similarly, the error at any element $i$ of $\mathbf{o}$ by LPA is given as:

$error_{LPA}^{i}=\mathbb{E\mid}\mathbf{o}\left[i\right]-\mathbf{c}\left[i\right]\mid=\mathbb{E}\mid Lap\left(\lambda\right)\mid=$
\begin{equation}\label{diff-error}
\sqrt{2\lambda}=\sqrt{2}\Delta\left(\mathbf{Q}\right)/\epsilon
\end{equation}

The higher error in output largely deviates results from the actual values; however, it would reduce the accuracy of the utility e.g., it would recommend more generic services. This LPA method results in high error due to the accumulating noise in output process \cite{fung2015combining}. Hence, to maximise utility, the authors in \cite{chan2011private, dwork2010differential} suggest reducing error by implementing a full binary input tree along with a logarithmic scale noise to the input. Another similar scheme \cite{fung2015combining} also implements differential privacy for strong location privacy.

\subsection{Evaluating Personalised Services}\label{personalised-services}
Recall that the output $\mathbf{o}$ from the process presented in Section \ref{profiling-diff} also consists of private profiling interests with differentially added noise, which is sent to the service marketplace, as shown in Figure \ref{system-process}. There are several ways to evaluate personalized contents and services \cite{isinkaye2015recommendation}. In our scenario, we evaluate personalised services based on \textit{similarity} metric for simplicity, whereas other techniques could also be used, e.g., content-based, collaborative, or hybrid filtering techniques. 

The service marketplace evaluates a similarity match between the keywords from the profiling interests (i.e., $\kappa_{\mathbf{o}}$ and from various services (i.e., $\kappa_{s}$). 

The similarity metric is evaluated based on the $tf \cdot idf$ metric \cite{manning2008introduction}.

Let $D'$ is the set of ads from all the advertisers wanting to advertise with an advertising agency. Each ad $ A{d_{ID}}$ is represented with a unique identifier $ID=1,..., D'$ and is characterised by $\kappa _{A{d_{ID}}}$ keywords. Let the function $f$ returns the set of keywords of an ad i.e. $f\left( {A{d_{ID}}} \right) = {\kappa _{A{d_{ID}}}}$ from the pre-defined list of ads and their corresponding keywords. These set of keywords are then used by another function $g$ for calculating an appropriate interest(s) by calculating the similarity between the ad's keywords \(\kappa _{A{d_{ID}}}\) and the list of keywords of individual interests i.e. \(\kappa _{k,l}\).


\[g\left( {{\kappa _{A{d_{ID}}}}} \right) = {\kappa _{k,l}}:sim\left( {{\kappa _{k,l}},{\kappa _{A{d_{ID}}}}} \right) = \]
\begin{equation}
\bigcup\limits_{\forall Ad{'_{ID}} \in D'} {\max \left( {sim\left( {{\kappa _{k,l}},{\kappa _{Ad{'_{ID}}}}} \right)} \right)}
\end{equation}

This similarity \(sim\left( {{\kappa _{k,l}},{\kappa _{A{d_{ID}}}}} \right)\) based on $tf \cdot idf$ is calculated as follows: Let $t$ be a particular keyword in \({{\kappa _{A{d_{ID}}}}}\) and $d$ be the set of keywords in \({\kappa _{k,l}}\); then the \textit{term frequency} (occurrences of a particular keyword $t$ in $d$) is $tf_{t,d}$. The \textit{inverse document frequency} of $t \in {\kappa _{A{d_{ID}}}}$ within $d$ is calculated as $idf_t= log \frac {N}{df_t}$, where $N$ is the collection of keywords of interests in \textit{Interest} categories and the $df_{t}$ is the \textit{document frequency} that is the number of \textit{Interest} categories that contain $t$. The $tf \cdot idf$ is then calculated as $tf \cdot id{f_{t,d}} = t{f_{t,d}} \times id{f_t}$. Thus the score for ${\kappa _{A{d_{ID}}}}$ can be calculated as:

\begin{equation}
score\left( {{\kappa _{A{d_{ID}}}},d} \right) = \sum\limits_{t \in {\kappa _{A{d_{ID}}}}} {tf \cdot id{f_{t,d}}}
\end{equation}

\subsection{Retrieving Personalised Services via PIR}\label{viaPIR}
A marketplace has several services offered to the customers. Let $\mathcal{DB}$ maintains relevant data related to services $s_{i,j}$ such as, description, keywords, link to services, links to associated objects (e.g., objects present on CDNs (content delivery networks)), associated JavaScript files etc. This information is structured as an $r\times s$ matrix with $r$ rows, represents as a block of the database that consists of $s$ words of $w$ bits each, detailed representation can be found in \cite{devet2012optimally}. We utilise a multi-server Hybrid-PIR scheme \cite{devet2014best} that further implements the IT-PIR \cite{gertner1998random} with CPIR \cite{micali2005optimal} that has the properties of both the schemes with lower communication cost. The servers in this setting are \textit{curious but not malicious} and do not collude to discover the query information. The aim of using this scheme in a multi-server environment is to privately opt-in and strongly conceal its interactions for particular services on the server side. 

A user wishes to privately retrieve $\beta^{th}$ record i.e., various relevant information related to a particular service $s_{i,j}$, from the database servers $\mathcal{DB}_{n}$. The client first derives the index of personalised services (as shown in Section \ref{personalised-services}), subsequent, it encodes a PIR query for those services against the $\mathcal{DB}$, and sends this query to the servers. Following, the servers evaluate the PIR response, encodes it, and responds back to the client. The client decodes the PIR response and obtains desired information about those services. Following, the client uses the desired service, which further contributes to the profiling process, as discussed in Sections \ref{profile-updating} and \ref{service-usage-profiling}. This is also depicted in Figure \ref{system-process}.


\subsection{Measuring Randomness/Uncertainty in Private Data}\label{evaluate-entropy}
Releasing user data helps extract accurate insights, enabling personalized services in the marketplace. However, as previously discussed, it also poses privacy risks, including identity and sensitive attribute disclosure, potentially revealing an individual's identity. The authors in \cite{sweeney2000simple} found that quasi-identifiers (e.g., job title, individual's interests, etc. that are not unique identifiers themselves) lead to identification of 87\% of individuals in the US. Hence, it is important to evaluate the susceptibility of each attribute in a user profile before publishing data to the Local Server or marketplace, as shown in Figure \ref{system-process}. The authors in \cite{majeed2020attribute} propose an algorithm that quantifies the susceptibility and entropy of attributes present in a user's dataset before releasing data. 

Let $D=N_{r,s},I_{r}^{\tau}$ is the dataset present on the LS that contain profiling interests with some probability distribution $P_{n}=\left(p_{1},\ldots,p_{n}\right)$, with $p_{i}\geq0 ;i=1,\ldots,n$ and $\underset{i=1}{\overset{n}{\sum}}p_{i}=1$, the entropy $H$ is given as follows:

\begin{equation}\label{privacy-loss}
L\left(I_{r},t\right)=H^{*}\left(I_{r}\right)-H\left(I_{r},t\right)
\end{equation}

\begin{equation}\label{entropy}
H\left(I_{r},t\right)=\underset{\forall i}{\overset{\mid I_{r}\mid}{\sum}}w_{i}\left(\underset{\forall i}{\sum}\left(-p_{i}log_{2}\left(p_{i}\right)\right)\right)
\end{equation}

The $ L\left(I_{r},t\right)$ represents the privacy loss at time $t$ when a subset of attributes of $ I_r$ might have been disclosed; the $ H^{*}\left(I_{r}\right)$ represents maximum entropy, whereas $ H\left(I_{r},t\right)$ is computed when the probability distribution of $ p_{i}$ is uniform, i.e., a violator may assign uniform probability distribution to each attribute when s(he) has no prior information about the values of each attribute. Similarly, $ H\left(I_{r},t\right)$ shows the entropy at time $t$ as shown in Eq. \ref{entropy}. In addition, the term $ w_{i}$ captures the relative privacy dominance/value of relative attributes e.g., $ w_{i}=1$ means that all the attributes in $I_r$ are equally important. 

\begin{figure}[h]
\begin{center}
\includegraphics[scale=0.35]{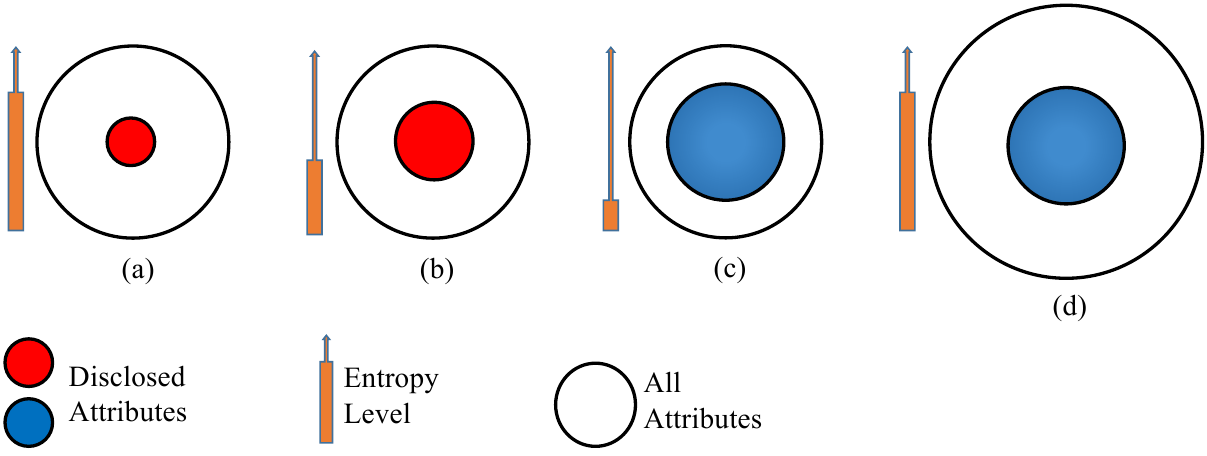}
\caption{System entropy: (a) when entropy reaches threshold (b) invoke \textit{data evaporation} to artificially increase entropy by controlled data distortion (c) invoke \textit{apoptosis} to destroy the private profiling attributes (d) entropy increases and disclosed attributes become less private due to e.g., more profiling attributes available now or private attributes become obsolete.}
\label{fig-entropy}
\end{center}
\end{figure}

Suppose that after time $t$, the violator figures out the state of the attributes within $ I_r$, which may allow her to know parts of the private attributes in $ I_r$, thenceforth, the system may decrease the entropy with attribute disclosure. For this, there are different options to decrease the system entropy, as shown in Figure \ref{fig-entropy}. E.g., \ref{fig-entropy} (a) when entropy reaches threshold then \ref{fig-entropy} (b) the \textit{data evaporation} can be invoked to artificially increase entropy by controlled data distortion, or \ref{fig-entropy} (c) when entropy drops to a certain low threshold then invoke \textit{apoptosis} process i.e., destroy the private profiling attributes and `re-evaluate user profile' (as also depicted in Figure \ref{system-process}) or \ref{fig-entropy} (d) when entropy increases as the entire set of attributes grows or when disclosed attributes become less valuable due to either private attributes become obsolete or private attributes dilute due to availability of more profiling attributes.

\section{Performance Measures}\label{performance-measures}
The section presents the applicability and implementation of critical components of the proposed model and further evaluates its performance using various evaluation metrics. We start with a case study where we present how to implement a service (in particular, the targeted behavioural advertising) evaluation metrics once the proposed model is implemented. We discuss the PoC implementation, on Android-based devices, of our proposed model that implements PIR components and local user profiling. 

\begin{figure}[h]
\begin{center}
\includegraphics[scale=0.35]{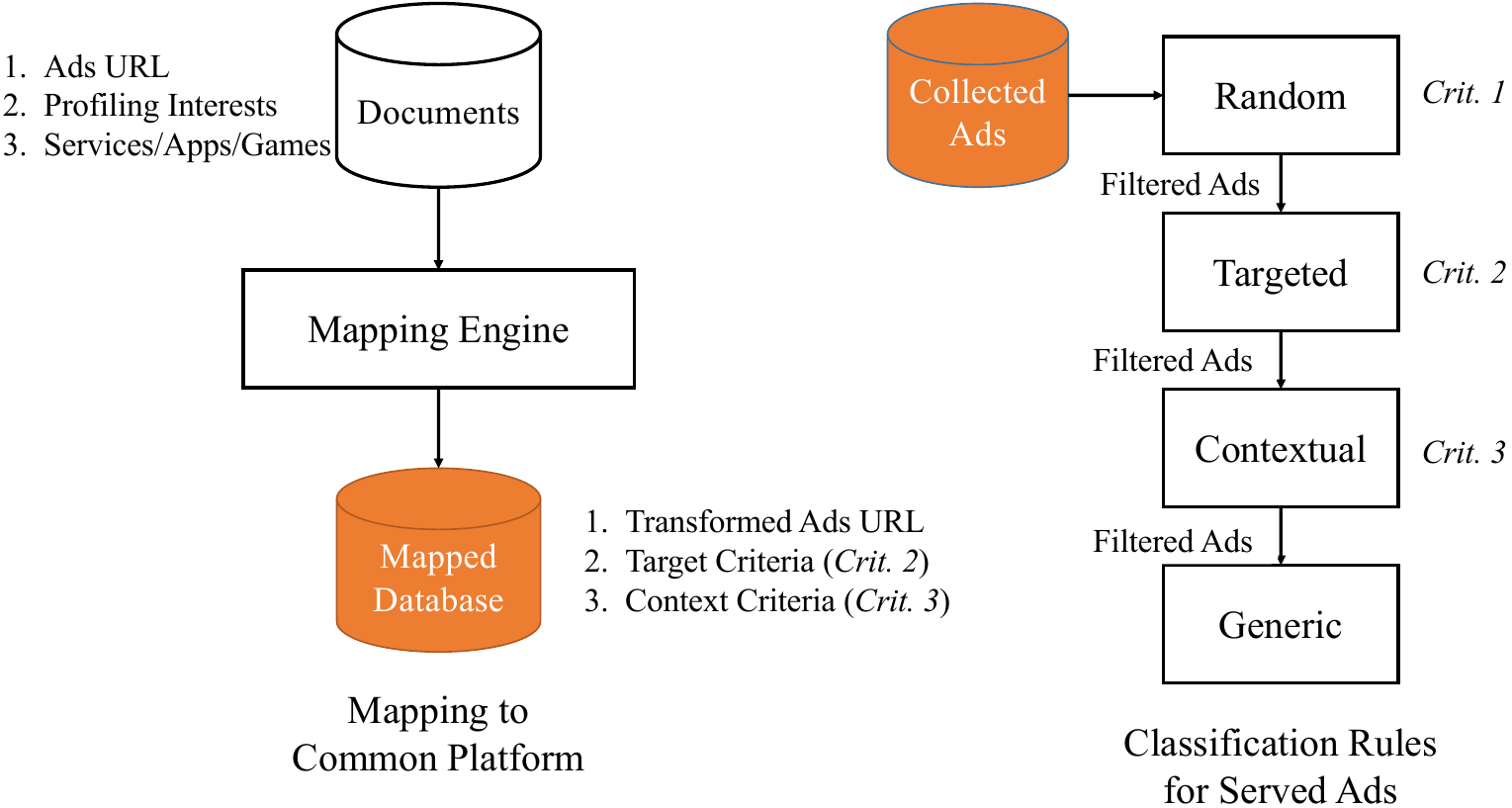}
\caption{Mapping rules for marketplace services, profiling interests, and ad-URLs (left). Classification rules for served ads is also shown (right).}
\label{mapping-rules}
\end{center}
\end{figure}

\subsection{A Case Study on Advertising Service}\label{performance-measures}
We study the advertising service to evaluate the impact of our proposed scheme and study its performance impact over the routine operations of a particular service within a marketplace. For this, we evaluate its impact over the behavioural advertising since the user profile is modified once the system produces a differentially private profile. Furthermore, we intend to carry out detailed performance analysis due to retrieval of targeted ads via PIR e.g., communication cost, processing overhead etc. Note that the ads (linked with a URL e.g., advertising \url{https://www.amazon.com/}) are served within the mobile apps as a banner that are either presented on the top/bottom of the screen or captures the entire device's screen. We formulate a uniform policy where these ad-URLs are mapped to a common platform, furthermore, define the \textit{context} and \textit{target} criteria. There are various online platforms, such as Cyren \footnote{\url{https://www.cyren.com/security-center/url-category-check-gate}}, or Url-Classification\footnote{\url{https://url-classification.io/}} that can be used to map URLs or evaluate \textit{context} and \textit{target} criteria. 

\paragraph{Mapping of Services and Profiling Interests} Let $\mathcal{C}$ is the set of categories defined by these platforms, comprise of the top level and their corresponding sub-categories and $c$ is a particular node that consists of hierarchy of categories from its own position to its root level. Thus, a particular node can be traced back to the root-level to find its ultimate category within these online platforms; such an example could be $c=$ \textit{\{Business/Accounting/Firms/Accountants\}}, with the root-level category of \textit{Business}. Hence, the mapping of a profiling interest categories is to find an appropriate node $c$ that is either on the top-level or any of the subsequent levels. We represent $C_s$ and $C_r$ respectively as the sets of transformed service's and profiling interest's categories, henceforth, will use these two sets respectively to compute the proportion of \textit{contextual} and \textit{targeted} ads. This mapping is also depicted in (the left side of) Figure \ref{mapping-rules} e.g., mapping (2) `Profiling Interests' to (2) `Target Criteria'.

\paragraph{Mapping of Served ad-URLs} Let an ad is represented as ${d_{r,s,q}}$ present under ${\mathbb{D}_{r,s}}$ as the set of ads served to different user profiles. ${\mathbb{D}_{r,s}} = \left\{ {{d_{r,s,q}}} \right\}$ where $r$ is specific \textit{interest profile}, $s$ is the apps categories, and $q$ represents the numbered ads. The mapping of an ad-URL can be done in the following steps: 1). if an ad-URL is pre-categorised by these platforms then the categorisation of the served ads is done directly i.e. ${E_{r,s}} = \left\{ {\left( {{d_{r,s,q}},c} \right):c = f\left( {{d_{r,s,q}}} \right)} \right\}$. The mapping $f$ returns the node $c$ by requesting mapping platforms under which a particular ad-URL is classified, such an example of categorisation of \url{www.google.com} is $c=$ \textit{\{Computers/Internet/Searching/Search Engines\}}. 
The set ${E_{r,s}}$ contains the mapped category $c$ associated with each ad ${d_{r,s,q}}$ in ${\mathbb{D}_{r,s}}$ and the ad ${d_{r,s,q}}$ itself. 2). Secondly, the reminders of the ads, not pre-categorised, are classified using a similarity match between the keywords derived from an ad-URL (if an ad-URL is not pre-categorised then the mapping platforms returns ads' keywords, which can also be automatically fetched from the source code of a URL) and the set of nodes $c$ in $C$. In this case, first the function $f$ returns the set of keywords of an ad-URL i.e. $f\left( {{d_{r,s,q}}} \right) = {\kappa_{r,s,q}}$. This set of keywords ${\kappa_{r,s,q}}$ is then used by function $g$:

\begin{equation}
{E_{r,s}} = \left\{ {\left( {{d_{r,s,q}},c} \right):c = g\left( {{\kappa_{r,s,q}}} \right)} \right\}
\end{equation}

\begin{equation}\label{eq-mapping}
g\left( {{\kappa_{r,s,q}}} \right) = c:sim\left( {{\kappa_{r,s,q}},c} \right) = \mathop {\max }\limits_{c' \in \,C} \left( {sim\left( {{\kappa_{r,s,q}},c'} \right)} \right)
\end{equation}

The $g$ calculates an appropriate node $c$ in $C$ by determining the similarity between the ad-URL's keywords ${\kappa_{r,s,q}}$ and the list of nodes $c$ in $C$, as shown in Eq. \ref{eq-mapping}, similar to the method presented in Section \ref{personalised-services}. This process is depicted in Figure \ref{mapping-rules} i.e., mapping (1) `Ads URL' to (1) `Transformed Ads URL'. Note that the served ads are either URLs of the sponsoring merchants or they are related to self-advertising i.e., the service marketplace advertises its own apps/services. Hence, the above method is used for calculating the ads' transformed categories when the ad-URL is related to the sponsoring merchants. Following, we use the sets ${E_{r,s}}$, $C_s$, and $C_r$ for the classification of served ads.


\paragraph{Rules for Classification of Served Ads} We define the ads targeting rules under which ads are classified into four different categories: \textit{random}, \textit{targeted}, \textit{contextual}, and \textit{generic} ads with the order of filtering for specific classes as shown in (the right part of) Figure \ref{mapping-rules}. We classify the \textit{random} ads as those that are served to all experimented devices (i.e., \textit{Crit. 1} as shown in Figure \ref{mapping-rules}) in a specified time period; detailed discussion over experimental setup is presented in Section \ref{exp-setup}. Hence, the set of \textit{random} ads $\delta _p^{ran}$ can be calculated as an intersection of ads i.e. the subscript $r$ in ${\mathbb{D}_{r,s}}$ is common for entire set of ads ${d_{r,s,q}}$ for all profiles $p \in P$:

\begin{equation}
\delta _p^{ran} = \bigcap\limits_{r \in \left[ {1,P} \right]} {{\mathbb{D}_{r,s}}}
\end{equation}

Note that these ads are removed from the pool of ads first, following, another class of the ads are calculated, such as \textit{targeted} ads, as shown in Figure \ref{mapping-rules}. The remaining filtered ads $\mathbb{D}_{r,s}^{\left( 1 \right)} = {\mathbb{D}_{r,s}} \setminus \delta _p^{ran}$ are passed through the \textit{target} criteria for calculating the \textit{targeted} ads. The \textit{targeted} ads $\delta _{r,s}^{tar}$, served corresponding to specific profiling interests $I_r$. These ads can be calculated as the intersection between the sets $E_{r,s}$ and $C_r$:

\begin{equation}
\delta _{r,s}^{tar} = \left\{ {{d_{r,s,q}} \in \mathbb{D}_{r,s}^{\left( 1 \right)}:f\left( {d_{r,s,q}} \right) \in \left( {{E_{r,s}} \cap {C_r}} \right)} \right\}
\end{equation}

Recall that mapping $f$ is two-folded i.e. it either directly fetches a node $c$ in $C$ of the pre-categorised URL or it determines $c$ by first fetching the set of keywords of an ad-URL and then calculating the similarity using $g$ between these keywords and the set of \texttt{Alexa} categories. Hence, \textit{targeted} ads are classified according to the matching criteria, by which an ad defined by the landing page URL has a high similarity with one or more interest categories in user profiles.

The reminders of the ads in $\mathbb{D}_{r,s}$ i.e. $\mathbb{D}_{r,s}^{\left( 2 \right)} = {\mathbb{D}_{r,s}} \setminus \left( {\delta _p^{ran} \cup \delta _{r,s}^{tar}} \right)$, are evaluated for \textit{contextual} ads $\delta _{r,s}^{con}$ according to the application \textit{context}. Hence, the \textit{contextual} ads are classified according to (again) a match between the \texttt{Alexa} transformed apps' categories $C_s$ and the keywords from the ad-URLs.

\begin{equation}
\delta _{r,s}^{con} = \left\{ {{d_{r,s,q}} \in \mathbb{D}_{r,s}^{\left( 2 \right)}:f\left( {d_{r,s,q}} \right) \in \left( {{E_{r,s}} \cap {C_s}} \right)} \right\}
\end{equation}

Finally, the ads that cannot be classified in any of the above methods are classified as \textit{generic} ads $\delta _p^{gen}$:

\begin{equation}
\delta _p^{gen} = {\mathbb{D}_{r,s}} \setminus \left( {\delta _p^{ran} \cup \delta _{r,s}^{tar} \cup \delta _p^{con}} \right)
\end{equation}

Note that other classification approaches could be used, for instance, one can deem ads uniquely served to some specific profiles, as \textit{targeted} ads, and consider others as \textit{generic}. 

\begin{figure}[h]
\begin{center}
\includegraphics[scale=0.35]{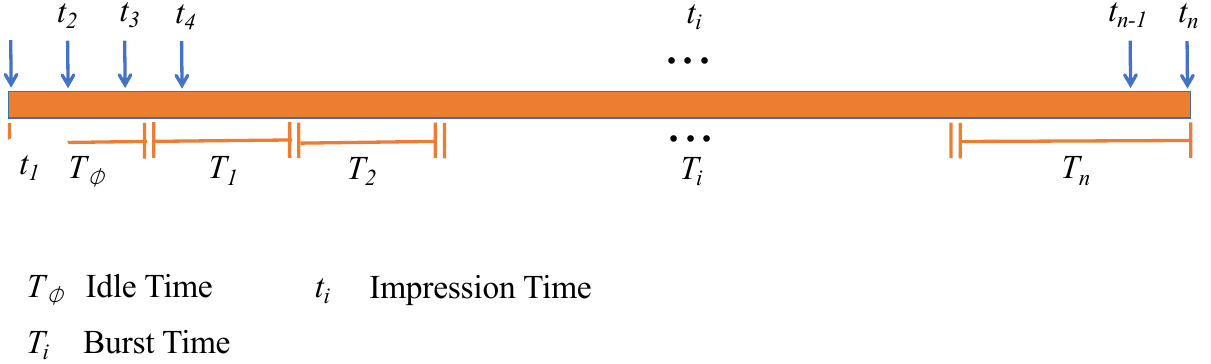}
\caption{Various parameters for \textit{idle} time $T_\emptyset$, ads \textit{impression} time (${t_{i + 1}} - {t_i}$) and \textit{burst} time $T_i$.}
\label{figure-adTimers}
\end{center}
\end{figure}

\paragraph{Additional Measures/Statistics} 
We evaluate various other parameters for ads analysis in an experiment, as depicted in Figure \ref{figure-adTimers}, in order to further evaluate ads-specific statistics e.g., the ads serving durations of various ads networks, consecutive ads from a specific ad network etc. These parameters are: The \textit{idle} time $T_\emptyset$ i.e., the time duration measured from the beginning of the experiment, where no ads are served; the ads \textit{impression} time (${t_{i + 1}} - {t_i}$) i.e., the time duration during which ads are shown on device's screen, where $t_i$ is the ad's \textit{inter-arrival} time. Note that the ad's \textit{inter-arrival} time is set by the application developers during apps registration in an App market such as, the ads are shown to mobile users after each 30sec, 45sec, 60sec etc. intervals, in their apps. In addition, we evaluate the \textit{burst} time $T_i$ (e.g., ${T_1} = \left( {{t_2} - {t_1}} \right) + \left( {{t_3} - {t_2}} \right)$, which is the time duration during which consecutive ads from a specific ad network e.g., Google AdMob, are served. 

\subsection{System Implementation}
\begin{figure*}[h]
\begin{center}
\includegraphics[scale=0.45]{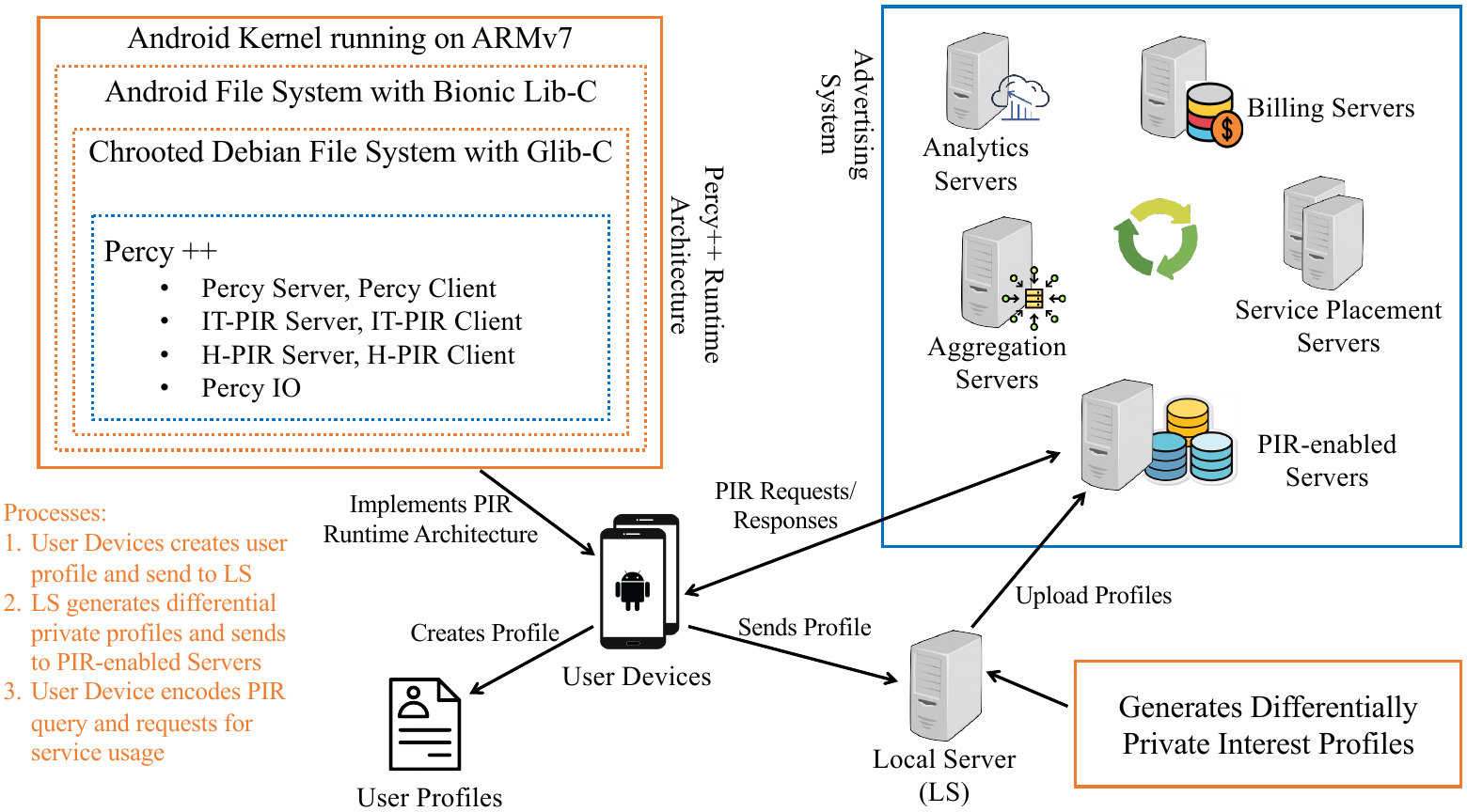}
\caption{Propose scheme implementation: Porting \texttt{Percy++} runtime architecture to Android device (left); Implementing Differential Privacy in LS; PIR query request/response (right).}
\label{percy-porting}
\end{center}
\end{figure*}

We implemented a PoC (Proof of Concept) app that implements critical components of the proposed scheme i.e., we develop local user profile based on user activities, implemented differential privacy over aggregated user profiles, and PIR implementation in Android-based device. The proposed system implementation is shown in Figure \ref{percy-porting}.

\subsubsection{User Profile Implementation} 
We use Android SDK and scripting to locally generate user profiles that represents the interests generated due to particular service usage (this also contains the installed apps from specific categories), the browsing history and searches, and the resultant interests profile. We evaluate the service usage by assigning weightages to individual service based on percentage of time a particular service (e.g., location service usage) is used\footnote{Usage time for a particular app can be calculated via \texttt{adb logcat -v threadtime} via Android SDK in order to find running time. The \texttt{threadtime} is used to display the date, invocation time, tag, priority, TID, and PID of a thread issuing the message.}. This also includes assigning the lowest time to those apps that are installed but not used including system installed apps. We consider the 24hrs period as the profile \textit{updating} and \textit{evolution} time to reflect service usage and user's behavioural pattern. 


\subsubsection{Porting Percy++ to Android Device for PIR Implementation} \label{percy-envmnt}

We run \texttt{Percy++}\footnote{A C++ version of various PIR implementations i.e., IT-PIR, CPIR, Hybrid-PIR, can be downloaded from \url{https://percy.sourceforge.net/download.php}. These protocols provide $t$-private, $v$-Byzantine-robust, $\tau$-independent, $k$-out-of-$l$ Private Information Retrieval.} on Android based device that required significant changes to support \texttt{Percy++}. Since we note that the CPU architecture of Android devices (i.e., commonly ARM: ARMv7 or armeabi) is different, the \texttt{Percy++} is mainly written for 32-bit/64-bit x86 CPU architectures. We implement the \texttt{Percy++} client architecture on Google Nexus 7, ARM Cortex-A9 Nvidia Tegra, 1GB RAM, 1.2 GHz quad-core. For this, we first install custom ROM CyanogenMod so as to get root access to Android device and to developer capabilities. We add support for Linux utilities that are not part of the Android devices i.e., the standard GNU C library\footnote{The Glibc found on most of the UNIX variant OS: \url{http://www.gnu.org/software/libc/}}. On the other hand, the Android OS is designed to use Bionic C\footnote{\url{https://android.googlesource.com/platform/bionic/}}, hence, we made this choice to enable high performing standard C library and its compatibility with the \texttt{Percy++} implementation. 

Figure \ref{percy-porting} shows the implementation of various components of our proposed architecture, specifically, the `\textit{Percy++} Runtime Architecture', the `Differential Privacy' component implemented on LS, the multi-server PIR implementation that is shown as part of the `Advertising System'. The `Chrooted Debian File System with Glib-C' component of `\textit{Percy++} Runtime Architecture' is a system call that enables to create a complete Ubuntu file system, which can run Debian environment on the Android Kernel ARM CPU support; they both share the same Linux Kernel. Similarly, the LS accumulates the \textit{interest Profiles} from individual users and associates similar profiles (for simplicity a 60\% to 100\% similar profiles might be grouped together) along with temporary IDs; note that we do not intend to send the permanent IDs e.g., similar to the `Android Advertising ID (AAID)\footnote{AAID: An anonymised string of numbers and letters generated for the device upon initial setup, e.g., `bk9384xs-p449-96ds-r132', \url{https://www.adjust.com/glossary/android-id-definition/ }}'. The user device encodes the PIR query and requests for services from the `PIR-enabled Servers' and gets the encoded response back from these servers; as discussed in Section \ref{viaPIR}. 

\section{Performance Evaluation}\label{evaluations}
We now present the experimental setup and detailed insights on evaluation of the proposed model.

\subsection{Experimental Setup}\label{exp-setup}

This section presents an overview of the experimental setup and the collected datasets. We perform two sets of experiments: 1). Experiments for collecting ad data by running mobile apps that focusses on investigating the extent of targeted advertising, specifically, for evaluating \textit{random}, \textit{targeted}, \textit{contextual}, and \textit{generic} ads) and evaluate the impact of proposed scheme (i.e., based on differentially private profiles) over targeted advertising. 2). Experiments for evaluating the applicability of implementing services based on requests/responses encoded with various PIR mechanisms. 

\paragraph{Experimental setup for ad collections} In particular, in first experimental setup, we are interested to study the influence of \textit{target} and \textit{context} for a particular user. There are many apps categories e.g., in Play Store\footnote{\url{https://play.google.com/}}, however, we randomly narrow down our selection to 9 app categories: `Business, Communication, Education, Entertainment, Games-Arcade \& Action, Health \& Fitness, Medical, Shopping and Sports'. We install the top 100 free apps from each of these categories and further narrow down their selection to run the 10 highest ranked apps from each category that receive ads. We run each app, from a specific category, for a period of 2.4hrs within a 24hrs period, over 9 phones in parallel, connected via Wi-Fi network. The aim of these experiments is to determine the \textit{profile establishment} i.e., \({K_s} \to {I_r}\), following, to achieve the \textit{stable} state of the profile, we continue running these apps for another 4 days, hence, with 10*24*9*5=10,800hrs (i.e., corresponds to 50 days of apps run over 9 phones) of total apps run. It is evident from our previous experimentations \cite{ullah2022joint} that the profile achieves a \textit{stable} profiling state where further activities of the same apps do not modify the profiling interests for a particular $I_r$. 

Following, for ad collection experiments during the \textit{stable} state, we run the same set of apps that were used for training during the \textit{profile establishment}. We run each experiment for a period of 24hrs, on 9 phones in parallel, which resulted in 10*24*9=2,160hrs of total apps run. We define an \textit{experimental instance} as the combination of an app usage in a specific apps category and a specific profile during our experiments for 24hrs e.g., an app run from `Business' category over `Education' profile for 2.4hrs. Hence, one \textit{experimental instance} is of 2.4hrs, where we collect data from 810 \textit{experimental instances}. We collect all traffic sent and received by the phones during the ads collection experiments using \textit{tcpdump}\footnote{\url{www.tcpdump.org}}. The apps run and data collection process is automated using the Android Debug Bridge\footnote{\url{developer.android.com/tools/help/adb.html}}, a tool that allows communication between a PC and a connected Android device. Although our setup is limited by the availability of phones, we consider that the chosen app categories and the corresponding apps within each category are a representative sample of the entire set of apps in Play Store. We argue that the results presented here can be generalised to other apps categories. Note that all our experiments were performed in one geographical region, furthermore, we acknowledge that the volume and diversity of the ad pool may vary from one region to another.

\paragraph{PIR experimental setup} This experimental setup simulates the exact runtime architecture for porting \textit{Percy++} to Android devices for PIR implementation, discussed in detail in Section \ref{percy-envmnt}. We use an Android phone to run the PIR client, with a 2.3GHz quad-core processor, 1GB RAM, running the 5.1.1 Lollipop OS to run the modified Percy++ client. We deploy the PIR server on a 3GHz Intel (R) Core (TM) i5-2320 with 32GB RAM. To implement multi-server IT--PIR and H--PIR, that require multiple servers, we run a minimum $l=3$ instances on a single server machine i.e., the minimum number of servers required to provide protection against $t=2$ colluding servers; we further vary number of servers from 3--6. In addition, we use the default \textit{Percy++} settings for \textit{word size} $w=10, 20$ bits respectively for IT--PIR and H--PIR, along with default recursive depth for H--PIR i.e., $\sqrt[d]{n}$, where $n$ is the total number of records in the database. 

In addition, we set other parameters for ads-database, such as, for the record (i.e., ad) size $s$ we take the a \textit{min}, \textit{max}, and \textit{avg} sizes of $12KB$, $20KB$, and $16KB$ respectively. These sizes of ads were taken based on our previous study and experiments \cite{ullah2022joint}, mainly composed 8--10 objects e.g., \texttt{JavaScript} files, images, other \texttt{HTML} objects etc., with an average of 30--35 request/response messages e.g., contacting \texttt{CDN} for downloading images. We argue that an $0.5GB$ of database size would adequately represent a good number of unique ads, however, we also experiment with higher range of database size of up to $100GB$. 

We measure the \textit{processing time} on both client and server (mainly the query encode/decode time and server processing time) with varying sizes of databases. Furthermore, we evaluate the \textit{communication cost}, in both directions, that is the bandwidth usage of ads traffic characterised by various components, such as query encode/decode sizes. We assess various PIR schemes via overheads introduced based on those two metrics, which may directly impact users with fixed data plan or the time it takes for landing ads on the device's screen. However, for the \textit{processing time}, we note that the \textit{ad refresh time} i.e., the inter-arrival time between two consecutive ads, varies between 20--60 seconds \cite{ullah2022joint}, hence, there might be sufficient time to request subsequent ads during the \textit{ad refresh time}. We carry out additional experiments with varying number of ads i.e., 1--8, are retrieved, which can be stored locally to subsequently present to the users, that might help save communication bandwidth and processing time. The practicality of the selected PIR scheme is assessed based on the amount of overhead it introduces to both metrics, as this will directly impact the user and could reduce the effectiveness of targeting (in the current ad system, there is a minimal delay between the ad request and delivery).

\subsection{Experimental Results}

Following, we present insights on the effect of differential privacy and percentage difference compared to the original set of ads, in addition, we present associated statistics, such as overall ads, unique ads and their frequency distribution, impression and idle time etc. Following, we present detailed investigation of various PIR schemes based on experimental setup discussed in previous section.

\begin{figure*}[t]
     \begin{center}
      \subfigure[Normal Vs. Differential privacy]{
            \includegraphics[width=0.5\columnwidth]{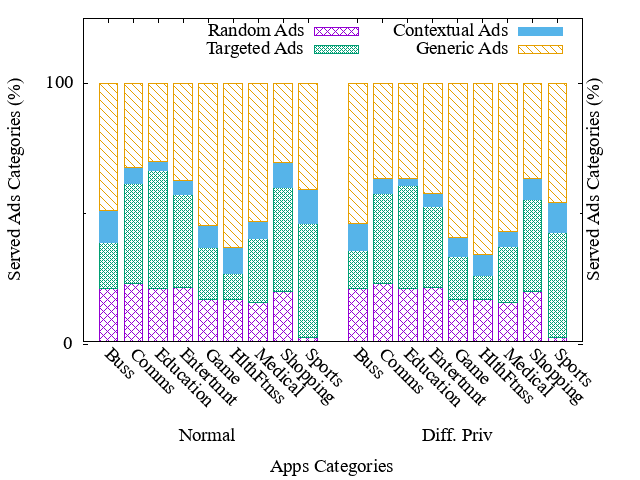}
        }
      \subfigure[Percentage Increase/Decrease]{
            \includegraphics[width=0.5\columnwidth]{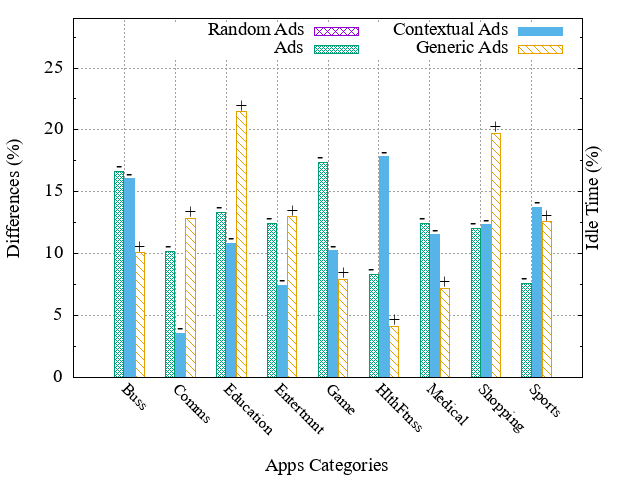}
        }
      \subfigure[Distribution of ads and time]{
            \includegraphics[width=0.5\columnwidth]{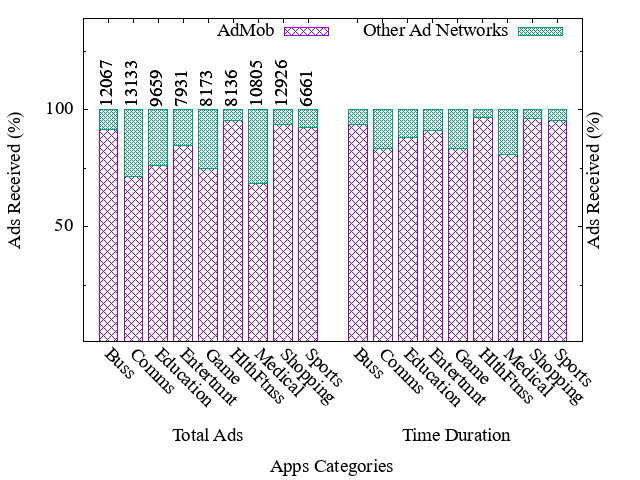}
        }
      \subfigure[Impression Vs. Idle Time]{
            \includegraphics[width=0.5\columnwidth]{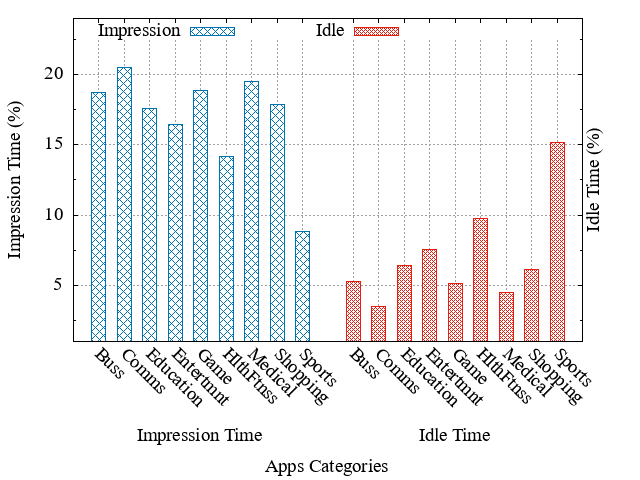}
        }
      \subfigure[Unique ads distributions]{
            \includegraphics[width=0.5\columnwidth]{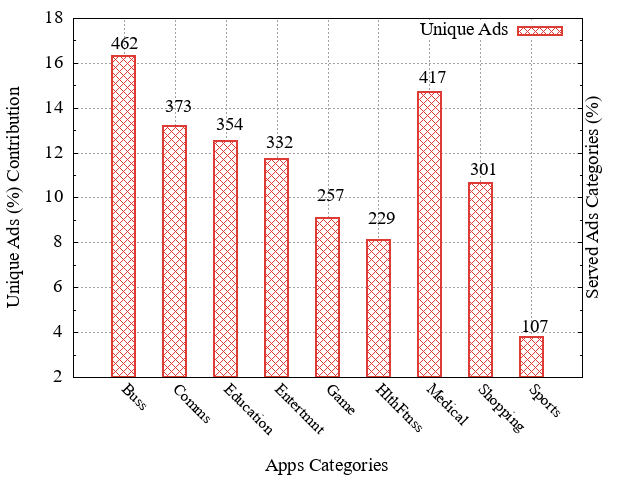}
        }
      \subfigure[Unique ads CDF]{
            \includegraphics[width=0.5\columnwidth]{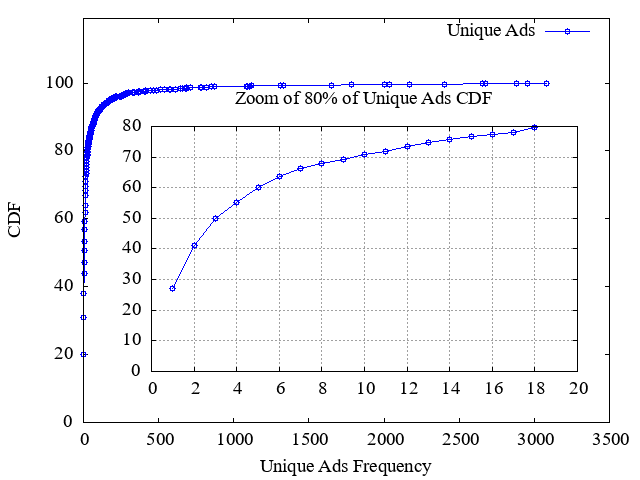}
        }
      \subfigure[Unique ads distribution among various frequency bins]{
            \includegraphics[width=0.5\columnwidth]{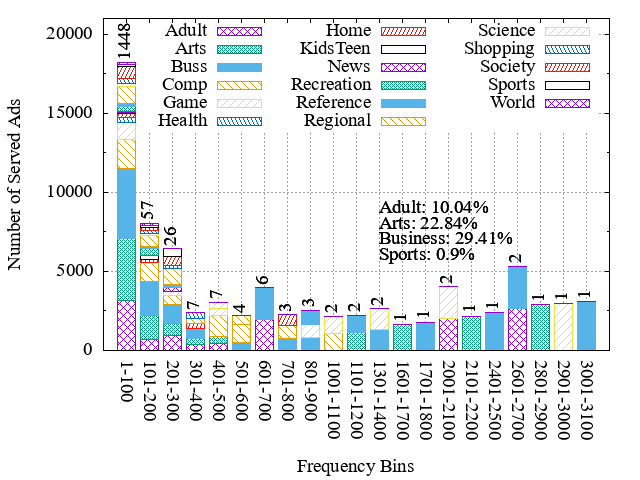}
        }
     \end{center}
    \caption{Various results evaluated for proposed strategy: (a) Impact of Differential Privacy over ads distributions, (b) Percentage increase/decrease calculated based on (a), (c) AdMob ads distribution and other networks ads with their \textit{impression} times, (d) Ads \textit{impression} and \textit{idle} time evaluated for various Apps categories, 
    (e) Unique ads distributions for various Apps categories, (f) CDF of Unique ads, and (g) Unique ads distribution among various frequency bins.}
     \label{figure-adsFigures}
\end{figure*}

\subsubsection{Impact of differential privacy over ads distribution}\label{ad-results}

\paragraph{Impact over ads distributions and percentage increase/decrease Figure \ref{figure-adsFigures} (a) and (b)}

We classify received ads into \textit{random}, \textit{targeted}, \textit{contextual} and \textit{generic} ads, collected via experimental setup as explained in Section \ref{exp-setup}. We calculate the \textit{random} ads with an overlap of 6hrs, however, we take another variant with 1hr overlap where we observe slightly lower number of ads; we do not present those results due to space limitations. Figure \ref{figure-adsFigures} (a) shows classification of received ads for various Apps profile (left) and the effect of differential privacy over the proportions of various classifications (right). In addition, the percentage differences of (various classes for averages and standard deviations) of are also presented in Table \ref{normal-diff-prv}. Note that the differential privacy has no effect on the \textit{random} ads, based on the evaluation criteria described in the Section \ref{performance-measures}.

It can be observed that the average proportion of \textit{targeted} ads is 30.06\%, which subsequently changes to 26.95\% with a difference of 3.65\% with the differential privacy. Similarly, these differences are 1.03\% and 4.68\% respectively for \textit{contextual} and \textit{generic} ads. We note that the differential privacy has slight effect over the percentage of received ads for various classes of received ads. Furthermore, as presented in Table \ref{normal-diff-prv}, we observe slight differences of St. Dev., both for received ads and differential privacy. In addition, Figure \ref{figure-adsFigures} (b) provides detailed evaluation of the percentage increase/decrease (i.e., evaluated for received ads and differential privacy) for various classes of received ads. We observe slight higher percentage change in the \textit{generic} ads for `Education' and `Sports', respectively a percentage increase of 21.46\% and 19.75\%, compared to the rest of the categories. Overall, we note slight differences percentage change, with an overall average of -12.03

\begin{table}
\begin{center}
\begin{tabular}{|c|c|c|c|c|}
\hline 
\multirow{2}{*}{Stats} & Random & Target & Context & Generic\tabularnewline
\cline{2-5} \cline{3-5} \cline{4-5} \cline{5-5} 
& \multicolumn{4}{c|}{Normal (\%)}\tabularnewline
\hline 
Avg. & 17.84 & 30.60 & 8.27 & 43.28\tabularnewline
\hline 
St.Dev. & 6.19 & 12.79 & 3.23 & 11.97\tabularnewline
\hline 
& \multicolumn{4}{c|}{Differential Privacy (\%)}\tabularnewline
\hline 
Avg. & 17.84 & 26.95 & 7.24 & 47.97\tabularnewline
\hline 
St.Dev. & 6.19 & 11.64 & 2.61 & 11.07\tabularnewline
\hline 
& \multicolumn{4}{c|}{Differences}\tabularnewline
\hline 
Avg. & 0.00 & -3.65 & -1.03 & 4.68\tabularnewline
\hline 
St.Dev. & 0.00 & -1.15 & -0.61 & -0.90\tabularnewline
\hline 
\end{tabular}\caption{Avg. and St. Dev. before and after applying differential privacy.}\label{normal-diff-prv}
\end{center}
\end{table}

Following, we present insights on other statistics about the received ads, such as the total ads, unique ads frequency distribution, and impression/idle/burst time of received ads. 

\paragraph{Additional statistics Figure \ref{figure-adsFigures}  (c) through (g)}

Figure \ref{figure-adsFigures} (c) shows ads distribution between AdMob and third-party (i.e., ad networks other than AdMob e.g., AdMarvel) ads along with the time duration for both advertising networks. The total number of ads for each apps category are also presented on the top of each bar, which sums up to 89,491 ads. For example, the `Business' category receive a total of 12,067 (distributed as 11,061 and 1,006 ads, respectively for AdMob and third-party ad networks) ads. These ads take the advertising time duration of 22.49hrs and 1.51hrs for AdMob and third-party ad networks, which corresponds to the 93.71\% and 1.61\% of the experiment time. We note that, on average, the AdMob consumes the 21.59hrs whereas the third-party ad networks take 2.41hrs of advertising airtime, with the standard deviation of 1.49hrs each for AdMob and third-party ad networks. Note that this advertising airtime consists of \textit{impression} and \textit{idle} time, as shown in Figure \ref{figure-adTimers}. Figure \ref{figure-adsFigures} (d) shows the \textit{impression} and \textit{idle} times for various apps categories; note that for simplicity we calculate these timers for the entire set of ads. We note slightly higher \textit{idle} time for the `Sports' category as opposed to other categories that has 6.03$\pm$1.96hrs. Likewise, the \textit{impression} time is 17.97$\pm$1.96hrs with 8.83hrs within the `Sports' category. 

We note that among 810 \textit{experimental instances}, 184 instances do not receive ads although those apps receive ads during the testing phase, as explained in Section \ref{exp-setup}. Possible reasons could be, for instance, the `Entertainment' profile attracts less number of ads when `Education' apps are run due to the influence of user profiles/interests, related to the presence, absence, or the contents of a user profile. To further support our argument, we additionally setup a phone with `Null' profile (i.e., opt-out of targeted ads) where we observe that it attracts the lowest number of total ads. In addition, 
we calculate the CDF \textit{burst} time (sec) of the advertising airtime, both for AdMob and third-party networks
Our analysis shows that consecutive ads from AdMob are more frequently served compared to combine the top 5 other Ad networks i.e. with the respective \textit{burst} time instances of 3,443 and 2,847 (combine). Furthermore, we observe that for some experiments this \textit{burst} time spans, for AdMob, over the entire duration of experiments where no ads are downloaded from any third-party ad networks.

Additionally, we evaluate the unique ads (i.e., an ad that uniquely appears in the entire sets of experiments) distribution (Figure \ref{figure-adsFigures} (e), unique ads for each apps category are also shown on the top of each bar) distributed among various apps categories. For these evaluations, we narrow down our selection to AdMob ads only since they are served with the highest propositions. We note that the `Business' and `Medical' apps categories receive the highest number of unique ads, respectively 462 and 417 ads, whereas the average number of ads are evaluated to 351$\pm$107 ads for all the tested apps categories. Note that these ads also consist the \textit{random} ads. Hence, the entire set of \textit{experimental instances} receive 1,578 unique ads. Furthermore, we evaluate the CDF of unique ads (i.e., the number of times every \textit{unique} ad is served, as shown in Figure \ref{figure-adsFigures} (f)) and explicitly evaluate the distribution of unique ads among various frequency bins (Figure \ref{figure-adsFigures} (g)) along with their ads' transformed categories. We note that 92\% of the ads are served within the range of 1--100 times compared to the rest 8\% that are served in the range 101--3100; e.g., one ad is served 3,081 times observed among various \textit{experimental instances}, which is also evident in Figure \ref{figure-adsFigures} (g). This figure also shows the Alexa transformed categories along with the received ads distributed among various (x-axis) frequency bins. We note that 10.04\% of the ads are categorised as the Alexa category `Adult', which is also shown in the figure together with other ads of higher proportions. The ads' high serving frequencies strongly suggest that ad networks could benefit commencing the use of caching mechanisms, such as the AdCache technique proposed in \cite{vallina2012breaking}, to efficiently utilise network resources; such as airtime, bandwidth etc., by caching ads of higher frequencies.

\subsubsection{Performance analysis for PIR}

\begin{table*}
\begin{center}
\begin{tabular}{|c|c|c|c|c|c|c|c|}
\hline 
\multirow{2}{*}{PIR schemes} & \multirow{2}{*}{DB size (GB)} & \multicolumn{3}{c|}{Communication bandwidth (MB)} & \multicolumn{3}{c|}{Processing time (sec)}\tabularnewline
\cline{3-8} \cline{4-8} \cline{5-8} \cline{6-8} \cline{7-8} \cline{8-8} 
 &  & Min & Max & Avg & Min & Max & Avg\tabularnewline
\hline 
\hline 
\multirow{12}{*}{IT-PIR} & 1 & 0.38 & 0.28 & 0.31 & 0.68  & 0.66  & 0.66\tabularnewline
\cline{2-8}
 & 2 & 0.71  & 0.48  & 0.56 & 1.35  & 1.32  & 1.33\tabularnewline
\cline{2-8}
 & 3 & 1.05  & 0.68  & 0.81 & 1.98  & 2.00  & 1.97\tabularnewline
\cline{2-8}
 & 4 & 1.38  & 0.88  & 1.06 & 2.66  & 2.66  & 2.63\tabularnewline
\cline{2-8}
 & 5 & 1.71  & 1.08  & 1.31 & 3.28  & 3.27  & 3.22\tabularnewline
\cline{2-8}
 & 6 & 2.05  & 1.28 & 1.56 & 3.93  & 3.88  & 3.93\tabularnewline
\cline{2-8}
 & 7 & 2.38  & 1.48  & 1.81 & 4.75  & 4.50  & 5.11\tabularnewline
\cline{2-8}
 & 8 & 2.71  & 1.68  & 2.06 & 5.26  & 5.19  & 5.26\tabularnewline
\cline{2-8}
 & 9 & 3.05  & 1.88  & 2.31 & 5.90  & 5.81  & 5.87\tabularnewline
\cline{2-8}
 & 10 & 3.38  & 2.08  & 2.56 & 6.53  & 6.42  & 6.45\tabularnewline
\cline{2-8}
 & \textit{Average} & 1.88  & 1.18  & 1.44 & 3.63  & 3.57  & 3.64\tabularnewline
\cline{2-8}
 & \textit{St. Dev} & 1.01  & 0.61  & 0.76 & 1.98  & 1.93  & 2.00\tabularnewline
\hline 
\hline 
\multirow{12}{*}{H-PIR} & 1 & 0.38  & 0.28  & 0.31 & 2.95  & 0.58 & 0.60\tabularnewline
\cline{2-8}
 & 2 & 0.96  & 0.48  & 0.56 & 3.82  & 1.17  & 1.17\tabularnewline
\cline{2-8}
 & 3 & 1.07  & 0.68  & 1.08 & 4.07  & 1.72  & \multicolumn{1}{c|}{1.73}\tabularnewline
\cline{2-8}
 & 4 & 1.18  & 1.20  & 1.17 & 4.61  & 2.30  & 2.31\tabularnewline
\cline{2-8}
 & 5 & 1.30  & 1.26  & 1.25 & 5.15  & 2.87  & 2.84\tabularnewline
\cline{2-8}
 & 6 & 1.39  & 1.33  & 1.36 & 5.77  & 3.39  & 3.41\tabularnewline
\cline{2-8}
 & 7 & 1.47  & 1.43 & 1.42 & 6.31  & 3.97  & 3.97\tabularnewline
\cline{2-8}
 & 8 & 1.57  & 1.48  & 1.48 & 6.87  & 4.53  & 4.50\tabularnewline
\cline{2-8}
 & 9 & 1.64  & 1.53  & 1.55 & 7.39  & 5.07  & 5.07\tabularnewline
\cline{2-8}
 & 10 & 1.70  & 1.58  & 1.63 & 7.92  & 5.66  & 5.61\tabularnewline
\cline{2-8}
 & \textit{Average} & 1.27  & 1.13  & 1.18 & 5.49  & 3.13  & 3.12\tabularnewline
\cline{2-8}
 & \textit{St. Dev} & 0.39  & 0.47  & 0.43 & 1.64  & 1.70  & 1.69\tabularnewline
\hline 
\end{tabular} \caption{Desktop client experiments: Communication bandwidth (MB) and total processing time (sec) for various sizes of ads (i.e., Min, Max, and Avg) and for ad database sizes of up to 10GB, evaluated for both IT-PIR and H-PIR schemes.}\label{pir-10-gb}
\end{center}
\end{table*}

\begin{figure*}[h]
     \begin{center}
      \subfigure[Query encode/decode size (MB)]{
            \includegraphics[width=0.5\columnwidth]{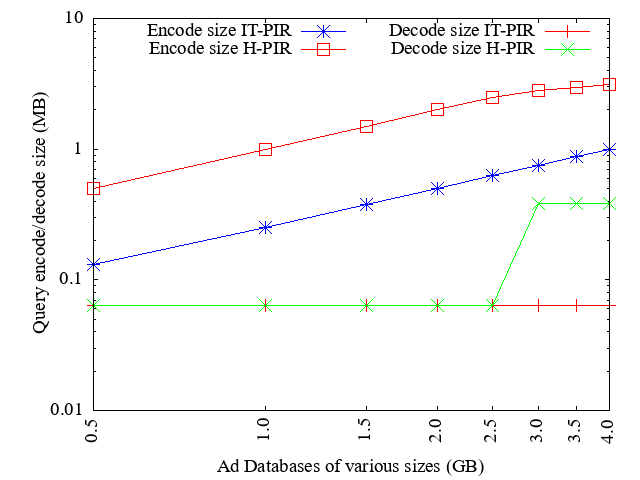}
        }
      \subfigure[Query encode/decode time (sec)]{
            \includegraphics[width=0.5\columnwidth]{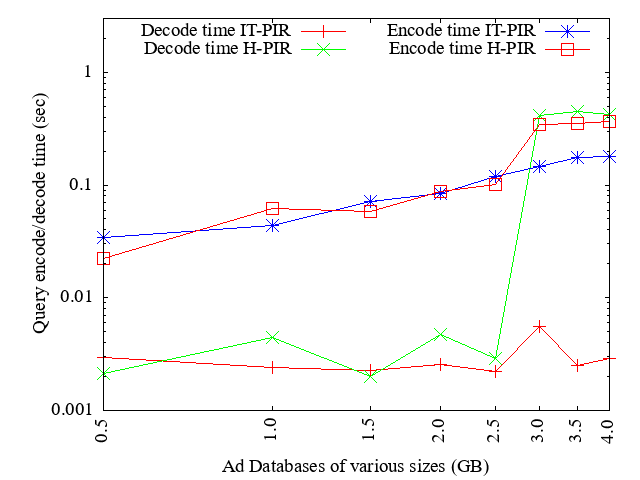}
        }
      \subfigure[Processing time (sec)]{
            \includegraphics[width=0.5\columnwidth]{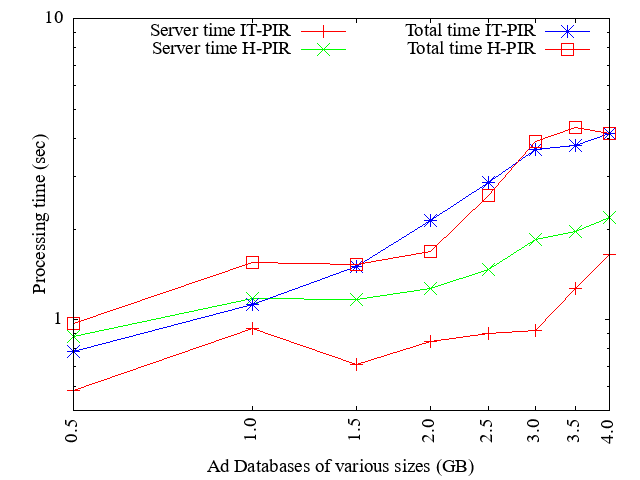}
        }
      \subfigure[Communication bandwidth (MB)]{
            \includegraphics[width=0.5\columnwidth]{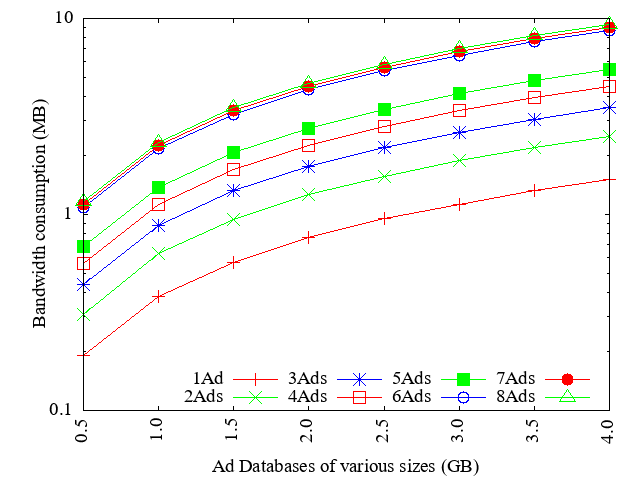}
        }
      \subfigure[Total processing time (sec)]{
            \includegraphics[width=0.5\columnwidth]{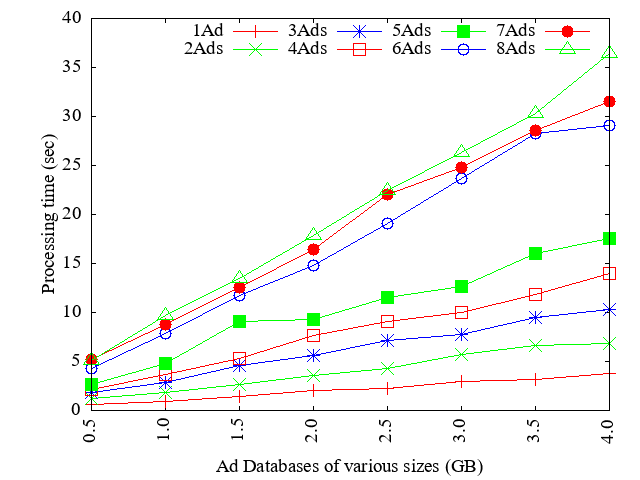}
        }
      \subfigure[Server processing time (sec)]{
            \includegraphics[width=0.5\columnwidth]{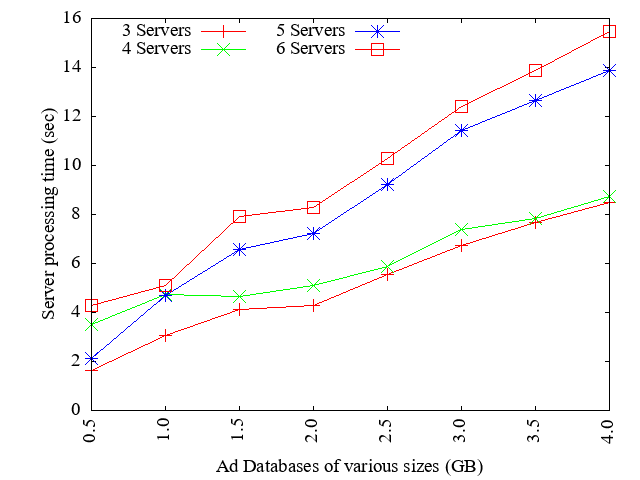}
        }
     \end{center}
     \caption{Android client experiments: (a) Query encode/decode size, (b) evaluation time, (c) total processing time for IT-PIR and H-PIR. (d–e) Impact of ads (1–8) on bandwidth and processing time. (f) Impact of server count ($l=3$–$6$) on server processing time.}
     \label{figure-pir-results}
  \end{figure*}

Recall that our main goal in this analysis is to evaluate the practicality of the PIR schemes in mobile advertising environment, specifically, to investigate the communication bandwidth and the processing time, since it will impact delay in advertising response and will affect users with fixed data plan. We note that, in the current ad system, there is minimal delay between the ad request, ad delivery, and response to the user action e.g., clicks, hence additional overhead by the PIR schemes might impact the effectiveness of ads targeting. We argue that the advertising systems can fully take the advantage of preserving user privacy during its communication with the ads systems, while privately retrieving various classes of ads along with their corresponding components from third-party networks/storages. 

Table \ref{pir-10-gb} shows the communication bandwidth and total processing time from $l=4$ server instances, ad sizes $s$ with \textit{min} $12KB$, \textit{max} $20KB$, \textit{avg} $16KB$, 1 ad request, and with database sizes of 1--10GB. These experiments are carried out on a desktop-based client and servers. We observe that the communication bandwidth (i.e., query encode/decode sizes) and processing time (note that this contains three time-overheads i.e., server processing time, query encode, and query decode time) increases with the varying sizes of databases; note this table for various ad-sizes of \textit{min}, \textit{max}, \textit{avg}. For instance, for the communication bandwidth, the \((avg,st.dev)\) pair for \textit{min}, \textit{max}, and \textit{avg} sizes for IT-PIR, respectively, are \((1.88,1.01)\)MB, \((1.18,0.61)\)MB, and \((1.44,0.76)\), compared to H-PIR, as \((1.27,0.39)\)MB, \((1.13,0.47)\)MB, and \((1.18,0.43)\)MB. Similar trend can also be observed for processing time (sec), for different sizes of database, with minor differences. As expected \cite{devet2014best}, the H-PIR provides the best results compared to IT-PIR scheme, since it automatically selects the recursive depth of CPIR \cite{melchor2007lattice} to optimise the communication bandwidth and improve the overall processing time.

Following, the Figure \ref{figure-pir-results} (a) through (c), shows detailed experiments for android-based client and desktop-based servers, $l=4$, both for IT-PIR and H-PIR schemes, however, we narrow down our selection to the \textit{avg} ad size, 1 ad request, and database sizes of 0.5--4GB. These experiments investigate the PIR query encode/decode sizes (MB) (that mainly contribute to the communication bandwidth), and the query encode/decode processing time (sec), the server and total processing time, which  contribute to the communication delay for delivering an ad to the client. We note that IT-PIR and H-PIR perform, similar to desktop-based client, quite efficiently on android-based client, to retrieve one ad from 0.5GB ad database size, with 0.78sec and 0.97sec delay, respectively, and 0.19MB and 0.56MB of communication data. Likewise, not surprisingly, large ad database sizes observe higher communication bandwidth and processing delay e.g., for a database of 4GB, it takes 4.14sec both for IT-PIR and H-PIR, and, respectively, a bandwidth consumption of 1.06MB and 3.51MB. Hence, for databases of all sizes, the observed average processing delays are 2.5sec and 2.59sec, respectively for IT-PIR and H-PIR, with \(st.dev\) of 1.3sec and 1.36sec, and bandwidth consumption of 0.63$\pm$0.31MB for IT-PIR and 2.23$\pm$1.10MB.

We also investigate H-PIR for the effect of varying number of ads i.e., between 1 and 8 ads, and varying databases of \textit{avg} ad sizes, desktop client, and $l=4$, in a single ad request and we compute the communication bandwidth (i.e., Figure \ref{figure-pir-results} (d)) and processing time (i.e., Figure \ref{figure-pir-results} (e)). We observe minute differences in the bandwidth consumption and delay as we increase the number of ads per ad-request from 1 to 8: e.g., on average, there is a 0.7MB increase in bandwidth consumption with an \(st.dev\) of 0.39MB for an ad database of 0.5GB. Moreover, \((avg,st.dev)\) pair for ad database of 4GB is \((5.57,3.09)\)MB. Hence, increasing database size from 0.5--4GB, we note an average bandwidth consumption of $3.13\pm1.71$MB and a \(st.dev\) of $1.74\pm0.95$MB. Similarly, as depicted in Figure \ref{figure-pir-results} (e), from 0.5--4GB of database and 1 to 8 ads per PIR request, we observe an average processing time of $10.89\pm5.59$sec and \(st.dev\) of $7.03\pm3.69$sec. We also investigate the effect of varying the number of servers from $l=3-6$ and mainly evaluate the server processing time; note that we also calculate other measures, however we only present how varying the number of servers impacts the server processing time for a single query request. As shown in Figure \ref{figure-pir-results} (f), the server processing time increases as the number of participating servers are increased for processing a single PIR request e.g., for ad database of 0.5GB, it takes 1.65sec with 3 H-PIR servers whereas it takes 4.27sec with 6 H-PIR servers to process 4 ads. Hence, on average, it takes 2.89sec with \(st.dev\) of 1.21sec as we increase the number of participating servers from 3 to 6. We note, ad database of 0.5--4GB and $l=3-6$, an average server processing time of $7.34\pm3.06$sec and \(st.dev\) of $2.21\pm0.95$sec. We argue that the processing delays with various PIR schemes are within the acceptable range of ad requests i.e., \textit{impression} time, as observed in the current ad system, which is also evident in our previous experiments presented in Section \ref{ad-results}.

\section{Discussion} \label{discussion} 


In our proposed solution, we address the threat scenario of active tracking of a service and user's internet activities by presenting a dual-ring protection scheme. This scheme utilises differential privacy to protect user profiling and private information retrieval to protect service dissemination. As all service providers implement their services over multiple servers and fulfill user requests simultaneously without collusion, private information retrieval is the most effective approach for private utilisation of distributed services. While anonymisation and randomisation are other potential privacy protection techniques, they do not fully meet the requirements of our threat model and the current service marketplace ecosystem.


Integrating the proposed framework with the existing advertising ecosystem is expected to be a straightforward process, as the PIR schemes are designed to be implemented over distributed servers without any collusion. The changes required would mainly be on the analytics servers, where the collection of personally identifiable information such as advertising ID, unique user identifier, ad impressions or clicks would need to be restricted, as we propose to move user profiling to the client side. However, the client side would require more changes, which can be embedded in the current implemented analytic SDK and libraries. These changes would include implementing user profiling on the client side, along with the incorporation of PIR schemes for encoding and decoding PIR queries. 


Although location privacy was not specifically addressed in our proposal, we suggest that it can be incorporated into the user profiling process as a demographic interest. Low-resolution GPS coordinates could be included in the user profile to enable advertisers and businesses to target potential customers based on their location. Alternatively, other privacy-preserving mechanisms such as Tor could be utilised alongside our proposed framework to protect location privacy.

\subsection{Service Usage Profile}\label{service-usage-profiling}
The service usage profiler\footnote{An example: Microsoft Usage profiler (Lifecycle Services, LCS) | Microsoft Learn, \url{ https://learn.microsoft.com/en-us/dynamicsax-2012/appuser-itpro/usage-profiler-lifecycle-services-lcs}} lets an organisation provide with a detailed summary of usage characteristics, such as transaction volumes, scheduling information, concurrent users etc. This will expose behaviour of their consumers who opt-in to specific services to the organisations \cite{ullah2014characterising, ullah2020privacy} or can be intercepted via connected networks \cite{tchen2014}. 

We represent \({U^t}\left( {{K_s}} \right)\) as the usage of the services at time slot $t$ for various services $K_s$; for simplicity, we do not consider behavioural usage for individual services. The service marketplace can disclose such interactions \cite{ullah2014characterising, ullah2020privacy}. Let \(\mathop {\lim }\limits_{t \to \infty } \sum\limits_{\tau = 1}^t {{U^\tau }\left( {{K_s}} \right)} \) denotes usage at different time slots, which varies during the time of the day. These updates can be incorporated in interest profile as: $I_r^{'t} = I_r^t + {U^{t}}\left( {{K_s}} \right)$.

Based on these profiling activities, the service marketplace can extract usage patterns to expose various information, such as peak hours of interaction, interactions with specific services, total peak concurrent users, etc. In addition, an adversary can also monitor employee's arrival/leave and work hours, or an organisation can expose an employee's productivity. We plan to collect data from a real environment, evaluate usage patterns, know the associated privacy risks, and provide comprehensive solutions for providing service usage profiling privacy. Hence, we do not consider this component in our current work and plan it for future work. 

\section{Conclusion}\label{conclusion}

The service marketplaces expose user privacy by aggregating and analysing the user's private sensitive profiling attributes and using it for service personalisation, targeted advertising, surveillance, or identity theft. This paper presents a dual-ring privacy protection mechanism that protects user privacy on the user and service sides. We aim to protect user privacy from various attacks, such as monitoring attacks, profile fingerprinting, stealing sensitive information, profile perturbation, selling sensitive information to third parties, exploiting profiling integrity, and evaluating targeted services. For this purpose, we use differential privacy to protect user profiling, with its additional protection via entropy and retrieving personalised services on differentially protected profiles via private information retrieval. Our extensive real-time data analysis shows that the proposed model provides better-profiling privacy by lowering the dominance of private attributes and provides private services with a minute impact over the targeted services for communication bandwidth consumption, delay, and target criteria. In addition, minute changes are required to integrate the proposed framework into the existing service marketplaces.

\printbibliography





\vfill
\end{document}